\documentclass{article}

\usepackage{arxiv}

\usepackage[utf8]{inputenc} 
\usepackage[T1]{fontenc}    
\usepackage{hyperref}       
\usepackage{url}            
\usepackage{booktabs}       
\usepackage{amsfonts}       
\usepackage{nicefrac}       
\usepackage{microtype}      
\usepackage{lipsum}
\usepackage{graphicx}
\graphicspath{ {./images/} }
\usepackage{caption, subcaption, float}
\usepackage{textcomp}
\usepackage{gensymb}
\usepackage[euler]{textgreek}
\usepackage{siunitx}
\usepackage{framed}
\usepackage{nomencl, multicol, multirow}
\usepackage{authblk}

\makenomenclature
\renewcommand*\nompreamble{\begin{multicols}{2}}
\renewcommand*\nompostamble{\end{multicols}}

\DeclareSIUnit\pixel{px}
\newcommand \dd[1]  { \,\textrm d{#1}}   
\title{Surface temperature and emissivity measurement for materials exposed to a flame through two-color IR-thermography}

\author[1]{Tanja Pelzmann}
\author[2]{Fabien Dupont}
\author[2]{Benjamin Saut\'e}
\author[1]{\'Etienne Robert}

\affil[1]{Polytechnique Montr\'eal, 2500 Chem. de Polytechnique, Montr\'eal, H3T 1J4, QC, Canada \authorcr \tt tanja.pelzmann@polymtl.ca}
\affil[2]{Telops Inc., 100-2600 St-Jean Baptiste Ave, Qu\'ebec, G2E6J5, QC, Canada \authorcr \tt fabien.dupont@telops.com}

\begin{document}
\maketitle
\begin{abstract}
Two-color (2C) pyrometry has long been used for flame temperature and soot concentration studies and is now becoming more widely used to measure surface temperatures of burning materials. With the obvious advantage of being a contact-free method that requires only minimal optical access, 2C pyrometry combined with high-speed acquisition is a promising diagnostic tool to obtain exceptional temporal and spatial resolution of thermally degrading samples.  However, its conceptual simplicity relies on a set of basic assumptions that when violated can result in large errors. In this work, we use an experimental configuration representative for fire resistance testing for aerospace and naval applications to analyze the impact of camera parameters and test setup on the accuracy of the surface temperature results obtained. Two types of fibre reinforced polymer composites and a steel plate are used to investigate material specific aspects that effect the measurements. An improved workflow for camera calibration is presented that takes the actual experimental setup into account. The temperature and emissivity mapping obtained trough in-situ IR measurements is compared against data acquired trough thermocouples and post-fire hemispherical directional reflectance measurements at room temperature. This comparison illustrates the necessity for proper post-processing and demonstrates that emissivity values obtained from pristine or burnt samples are not well suited to obtain accurate surface temperatures through conventional (single color) IR thermography. We also present a detailed error budget and suggestions for calibration measurements to keep the overall error well below \qty[]{50}{\degreeCelsius} in a temperature range from \qtyrange{400}{1200}{\degreeCelsius}.

\end{abstract}


\section{Introduction}\label{sec:intro}

Polymer matrix composites (PMC) are widely employed in the transportation sector and continue to replace traditional metal parts. Despite their versatile mechanical properties and the significant weight reduction they allow, important challenges must be addressed for applications at high temperature. Under these conditions, thermal degradation results in rapid changes in the material mechanical, thermal and optical properties. For aerospace or naval applications, these materials are subjected to certification testing where they are exposed to a calibrated heat source, typically a heating coil, a lamp, or a flame. For components located close to fire hazards, the tests involve a pilot flame fueled by either propane or oil. Predicting and understanding the thermal degradation mechanisms of PMCs requires precise knowledge of the material temperature, a quantity difficult to obtain experimentally under the conditions of certification fire tests. Optical techniques such infrared (IR) thermography can provide high spatial resolution, non-intrusive measurements but require careful calibration to yield accurate results, a time-consuming step not always implemented in practice. Here we introduce a novel two-color IR camera calibration approach that includes optical path modelling and geometrical corrections, drastically reducing the error budget associated with optical thermography, thus bridging the gap with point-based thermocouple (TC) measurements.

Regardless of the heat source used in fire testing, radiative heat transfer towards the sample is the main process that initiates and drives the PMCs thermal degradation \cite{Lautenberger_Fernandez-Pello_2009, Baukal_Gebhart_1996, Baukal_Gebhart_1995, Walter2006}. The specific experimental conditions affects both how the material degrades and how surface temperature can be measured by IR cameras. Two important factors are hence the spectral dependency of the sample absorptivity with respect to the heat source and the spatial heterogeneity in the heat source itself. Depending on the type of test apparatus, the latter may be less pronounced (cone calorimeter) or significant (high temperature lamp). The case of samples that ignite is even more complex as the surface flame is a secondary source of radiation and the emission pattern as a whole is affected. Numerical models to predict the thermal degradation heavily rely on sample emissivity and absorptivity that determine the energy distribution and heat loss \cite{Lautenberger_Fernandez-Pello_2009, Bal_Rein_2013}. Stoliarov et al. \cite{Stoliarov_Crowley_Lyon_Linteris_2009} reported uncertainties of \qty{\pm 50}{\percent} for the absorption coefficient and \qty{\pm 20}{\percent} for the reflectivity. This may be partly attributed to the design of their model, where all radiative properties are kept constant, although most of the material parameters are temperature dependent and change as it degrades. Studies on the surface reflectance and absorptance of several materials \cite{Forsth_Roos_2011}, polymers in particular \cite{Hallman_Welker_Sliepcevich_1974}, show that the spectral characteristics of the heat source, not only its nominal temperature, affect how much heat is actually absorbed by the exposed sample. The flame radiation of hydrocarbon fuels is located in the infrared spectrum \cite{Hood_1966}, where most polymers commonly have a high absorptance with little variation. However, sample emissivity and absorptance might still change as the material decomposes \cite{Boulet_Brissinger_Collin_Acem_Parent_2015}.  

\begin{table}[t]   
\begin{framed}
\nomenclature{PMC}{Polymer matrix composite}
\nomenclature{CFRP}{Carbon fibre reinforced polymer}
\nomenclature{\textepsilon}{Emissivity [-]}
\nomenclature{IR}{Infrared}
\nomenclature{MS}{Multi-spectral}
\nomenclature{\textPhi}{Equivalence ratio [-]}
\nomenclature{TC}{Thermocouples}
\nomenclature{IBR}{In-band radiance}
\nomenclature{NETD}{Noise equivalent temperature difference}
\nomenclature{RT}{Radiometric temperature [K]}
\nomenclature{FPA}{Focal plane area}
\nomenclature{AEC}{Automatic exposure control}
\nomenclature{ROI}{Region of interest}
\nomenclature{HDR}{Hemispherical Directional reflectometer}
\nomenclature{TES}{Temperature-emissivity separation}
\nomenclature{2C}{Two-color}
\nomenclature{NEdIBR}{Noise equivalent differential IBR}
\nomenclature{FTIR}{Fourier transform infrared spectroscopy}
\nomenclature{PFA}{Perfluoralkoxy (Telfon\textsuperscript{\textregistered})}
\nomenclature{SS}{Stainless steel}
\nomenclature{GF}{Glass fibre}
\nomenclature{1C}{Single color}
\nomenclature{OD}{Optical density}
\nomenclature{FOV}{Field of view}
\nomenclature{IA}{Illuminated area}
\printnomenclature
\end{framed}
\end{table}

The radiation absorbed by a sample causes its temperature to change, and the measurement of this change is essential to understand the outcome of many fire test. In large-scale certification setups, this measurement is commonly limited to a few thermocouples (TCs) at defined positions, allowing limited insights \cite{Apte2006} and coming with their own experimental challenges \cite{Shannon_Butler_2003}. Meanwhile, several bench-scale setups have been developed \cite{Tranchard_2015, Acem_Brissinger_Collin_Parent_Boulet_Quach_Batiot_Richard_Rogaume_2017, Sanchez-Carballido2018, Seggewiss_2011, Melendez2010, Rippe_Lattimer_2015, Bearinger_Hodges_Yang_Rippe_Lattimer_2020} that monitor the spatial and temporal temperature evolution through IR thermography at the sample back face, i.e. away from the heat source. The sample temperature is then obtained from an inversion of Plank’s law, which requires the sample emissivity as input.  Since this value is inherently difficult to assess experimentally, it is practical to assume it constant over the whole test period, taken either from the literature or measured before or after testing. However, the spectral emissivity of most materials varies with temperature, wavelength, and surface conditions. Thermal degradation affects all three variables and therefore the apparent temperature obtained by single-wavelength pyrometry may be seriously in error \cite{Reynolds_1964}. In several reports dealing specifically with PMCs \cite{Sanchez-Carballido2017, Schuhler_Chaudhary_Vieille_Coppalle_2021, Acem_Brissinger_Collin_Parent_Boulet_Quach_Batiot_Richard_Rogaume_2017, Melendez2010, Li_Gong_Stoliarov_2014} these significant contributions to the radiometric model are neglected, leading to measurements values being reported as overly precise. A potential approach to overcome this problem is two-color (2C) – or ratio – pyrometry. This technique is commonly used to measure soot concentration \cite{Reggeti_Agrawal_Bittle_2019, Zhang_Dai_Lu_Wu_2016}, temperature evolution in turbulent flames \cite{Yu_Bauer_Huber_Will_Cai_2021}, or for flame-wall impingement studies \cite{Arakawa_Saito_Gruver_1993, Chan_Fattah_2019, Aphale_DesJardin_2019}. The basic idea is to measure the sample radiance through two different narrow spectral bands and then use the signal ratio to calculate the object temperature \cite{Hunter_Allemand_Eagar_1986, Hunter1984, Inagaki_Okamoto_Fan_1994}. For thermally degrading materials such as PMCs, 2C IR pyrometry has not yet been implemented to yield precise assessment of the surface temperature with high spatial and temporal resolution. Moreover, the question of radiometric precision in one and two color pyrometry warrants a careful review in the context of fire testing.  

In this paper, we present a rigorous 2C pyrometry framework, covering hardware selection, camera calibration and post-processing, and implement it to measure the backside temperature of thermally degrading materials. The experimental campaign covers metallic samples and PMCs commonly used in aerospace and transportation applications. We discuss the potential sources of noise and error associated with the IR acquisition for rapidly degrading materials such as PMCs. We also develop a radiation model and assess the validity of the assumptions underlying 2C pyrometry and how they affect the accuracy of the surface temperature and emissivity measurements. This multi-step thermography image processing scheme offers a comprehensive approach to accurately study rapidly degrading materials under intense flame attack. 


\section{Material and methods}

\subsection{Sample panels}

Two types of carbon fibre and one type of glass fibre reinforced polymer (CFRP and GFRP, respectively) composite panels were tested, along with stainless-steel samples, to cover effects of emissivity and thermal conductivity. The two SAE 304 stainless steel panels measured \qtyproduct{305 x 305}{\milli\metre}, with a thickness of \qty{1.6}{\milli\metre}. Concentric circles with \qty{10}{\milli\metre} radial spacing were marked onto the backside, as reference for welding thermocouples (TC). All metal panels were used in the condition they were received in and do not display significant surface wear. Two CFRP panels with phenolic CYCOM 2400-1M resin and T650 carbon fibers (8 harness satin) were fabricated with four plies in [0/+45/-45/0]$_{\text{S}}$ orientation. Type K TC with a diameter of \qty{0.08}{\milli\metre}, coated with a thin layer PFA insulation to prevent electrical conduction with the material, were embedded in-between plies, as well as on each face to measure the temperature evolution throughout the sample thickness. The panels measured \qtyproduct{65 x 300}{\milli\metre} and were \qty{1.65}{\milli\metre} thick. CFRP samples were also manufactured with epoxy resin as \qtyproduct{25.4 x 305}{\milli\metre} wide coupon strips and used for flame calibration and some verification testing. A glass-fibre panel with phenolic resin (GF-Ph), \qtyproduct{35 x 300}{\milli\metre}, was used for emissivity measurements.

\subsection{Fire test configuration}\label{sec:testsetup}
\begin{figure}[tb]
    \centering
    \begin{subfigure}[c]{0.45\textwidth}
    \centering
    \includegraphics[height=45mm]{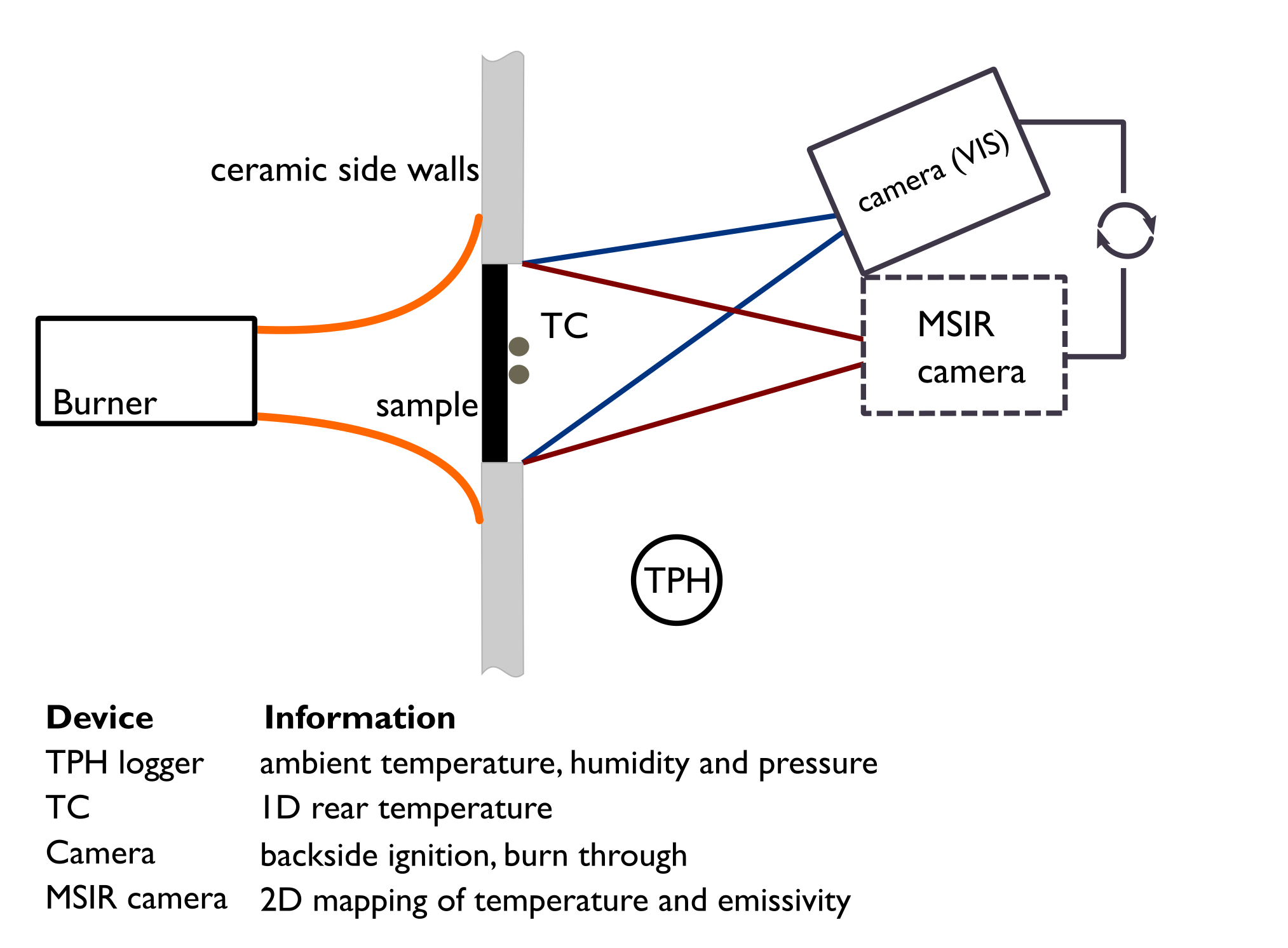}
    \caption{}\label{fig:setup0}
    \end{subfigure}
    \begin{subfigure}[c]{0.45\textwidth}
    \centering
    \includegraphics[height=45mm]{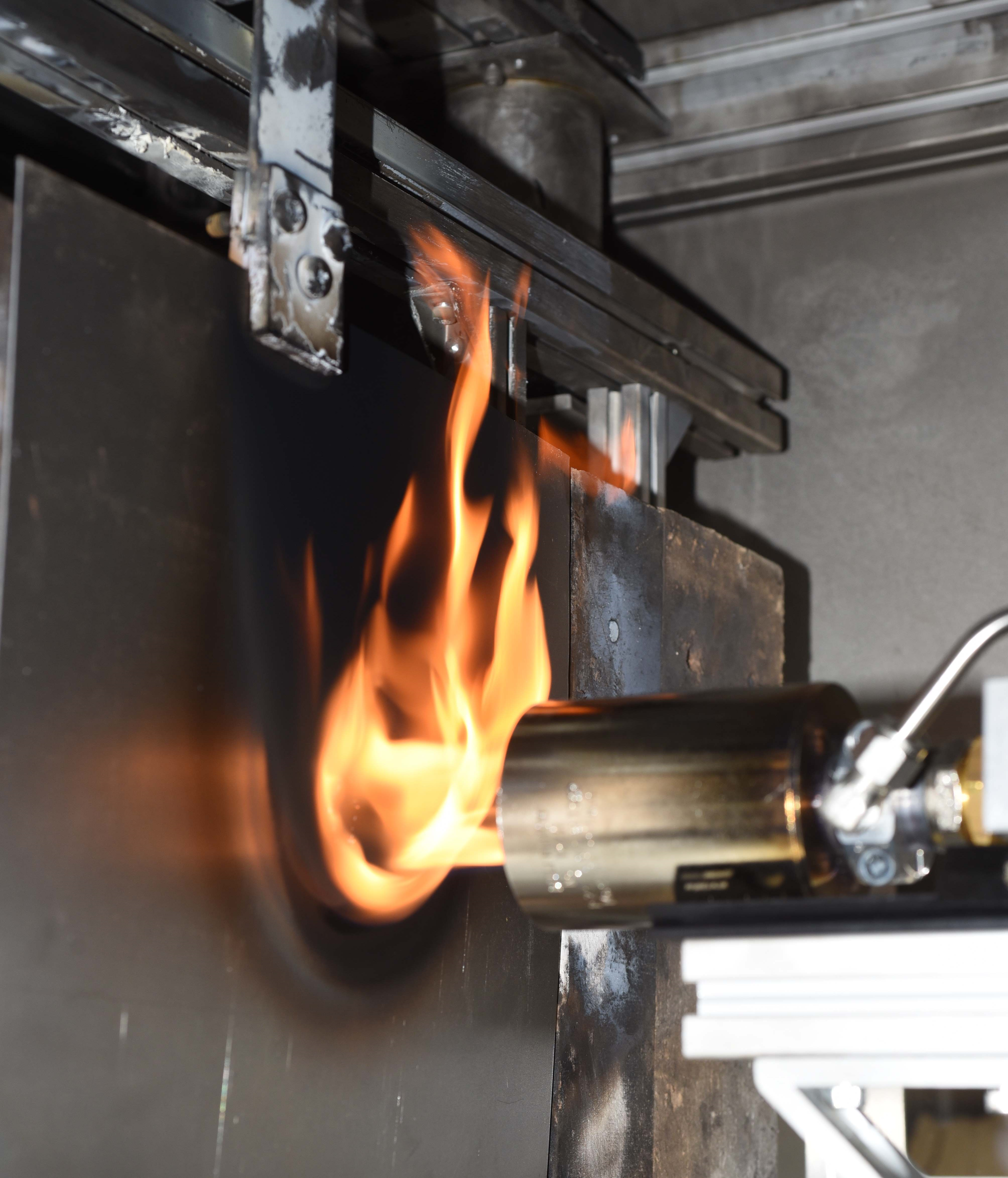}
    \caption{}\label{fig:flame}
    \end{subfigure}
    \caption{a) Schematic depiction of the experimental setup. The test rig is comprised of the following main parts: the exchangeable burner, the sample mounting grips and ceramic shields, exhaust and air entries and optical ports on four sides. b) Photograph of the propane flame impinging on a steel sample.}
    \label{fig:setup}
\end{figure}

The top view of the experimental configuration used to expose the material samples to open flames is shown in Fig.~\ref{fig:setup}. It is representative of fire certification tests where a vertically mounted sample is exposed to a horizontally impinging flame \cite{ac20135, iso2685}. An environmental chamber connected to a fume hood, with a volume of approximately one cubic meter, houses the sample mount and the burner. Samples with an area smaller than the flame cross section were shielded from both sides with insulating ceramic fibre boards (CeraMaterials), with a recessed step, to prevent the flame from wrapping around the specimen while maintaining the exposed surface uniform, as visible in Fig.~\ref{fig:setup}. The coupons are clamped on the bottom and top to vertically align the center of the sample with the flame. The mounting frame can be moved horizontally to control the distance to the burner exit. The chamber provides optical view ports on all four sides, with the main visualizations carried out from the back side using IR and visible cameras. A DSLR camera (Nikon D750) is also used to provide complementary imaging. 

\subsection{Gas and oil burner}\label{sec:burner}

A laboratory-scale kerosene burner \cite{Beland_Sc_2018} was used to generate conditions representative of the large-scale testing of aerospace components \cite{FAA}. Jet fuel-A is atomized through a \ang{60} hollow-cone bypass nozzle (Delavan 33769-2 with adapter no. 17147). Adjusting the inlet and bypass pressure allows the desired fuel flow rate, and hence burner power, to be attained. The oxidizer is laboratory compressed air, regulated with an Omega FMA5400 flow controller, allowing precise control over the flame chemistry. The air passes through a stream stator with 25 blades at \ang{15} orientation to enhance mixing with the fuel droplets and increase the evaporation rate. This swirling flame configuration was developed to mimic large-scale burners, such as the NexGen burner \cite{Ochs_2008, Kao_Tambe_Ochs_Summer_Jeng_2017}, prescribed in certification testing standards \cite{ac20135, iso2685}. A commercial off-the-shelf gas blow torch (Bullfinch No.1270) was used to generate propane flames, motivated by its use in the literature for similar fire resistance investigations \cite{Tranchard_2015, Gibson_Torres_Browne_Feih_Mouritz_2010, Horold_Schartel_Trappe_Korzen_Naumann_2012}. The burner has been modified with a collar around the air inlet to control the oxidizer amount and therefore the flame chemistry. 

\subsection{Temperature, emissivity, and high-speed infrared thermography measurements}

Type K TC were used to measure the sample temperature, with the probe design adapted to the sample type and test setup. These measurements served as second independent measurements for validation and comparison against IR thermography. Probes that measured the temperature on the sample back (cold) face were not in direct contact with the ﬂame and thus needed no protective coating to prevent catalytic reactions or surface degradation. For the steel samples, sturdy type K thermocouples (\qty{1.628}{\milli\metre} in diameter) were welded at different radial distance from the sample centre to ensure proper contact throughout the tests. The radius was increased in steps of \qty{10}{\mm}. Most of the composite samples had thin (\qty{0.08}{\milli\metre}) PFA-coated type K TC placed in-between each ply, with their beads slightly shifted relative to their neighbours, allowing the temperature gradient throughout the sample thickness to be measured without inducing local deformation. All thermocouples, except for those irreversibly integrated in composite panels, were tested prior to use with a dry-well (Fluke Calibration 9140). 

\begin{figure}[ht]
\centering
  \begin{subfigure}[t]{.45\textwidth}
    \includegraphics[width=\textwidth]{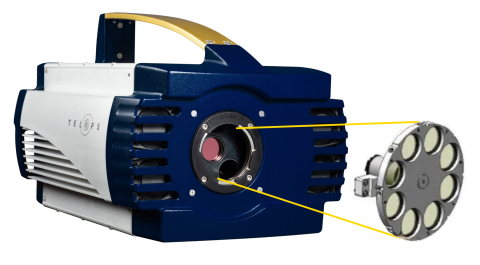}
    \caption{}
    \label{fig:msir}
  \end{subfigure}
   \begin{subfigure}[t]{.45\textwidth}
    \includegraphics[width=.9\textwidth]{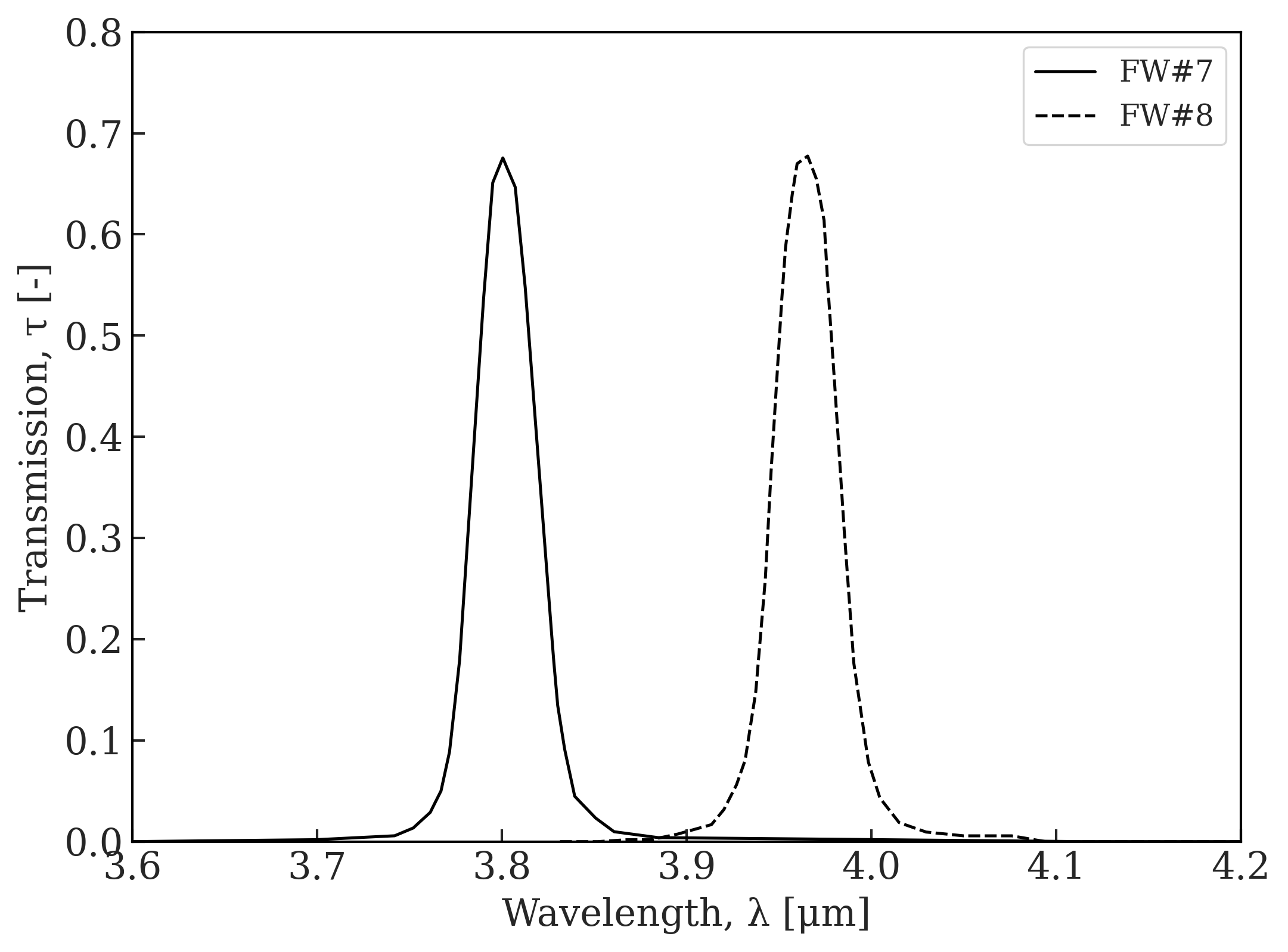} 
    \caption{}
    \label{fig:filter}
  \end{subfigure}
  \caption{(a) The Telops MSIR camera has a fast-rotating eight-position filter wheel. (b) Transmittance curves for the through-flame filters (cf. to Tab.~\ref{tab:camconfig}) provided by the camera manufacturer: NP-3800-040 (FW\#7) and NB-3950-040 (FW\#8).}
  \label{fig:cam}
    \end{figure}

The multispectral (MS)-IR camera (Telops FAST-M350) was conﬁgured with a \qty{50}{\milli\metre} lens and an eight-position fast-rotating filter wheel to record diﬀerent wavelength bands (Fig.~\ref{fig:cam}). The focal plane area (FPA) of the indium antimonide (InSb) detector is \qtyproduct{640 x 512}{\pixel} with a detector pitch of \qty{15}{\micro\metre}. The camera numerical aperture is F/3. The mid-wave IR filter configuration is listed in Tab.~\ref{tab:camconfig}. The optical density (OD) indicates the amount of energy blocked by a given filter. In the case of the neutral density filters in position FW\#2 and FW\#3, the transmission is evenly reduced across their spectral range to \qty{2.5}{\percent} and \qty{0.1}{\percent}, respectively. The filters FW\#7 and FW\#8 are so-called 'through-flame' filters because they limit contributions from combustion gases and ensure that the radiance contribution received by the sensor is dominated by the contribution from the hot surface. Software-defined automatic exposure control (AEC) was used to ensure proper exposition during strongly varying conditions and fast recording rates. The MS-IR camera was positioned perpendicular to the sample backside and the windows of the test chamber were unmounted to limit ray path obstruction. Radiance calibration measurements were performed with a Fluke Calibration 4181 black body. 

The emissivity of the samples is expected to change as the samples thermally degrade and their surface alters. To compare the initial and final emissivities obtained through IR measurements against alternative techniques, CFRP and GFRP samples have been analysed in their pristine state and post-fire. The absorptivity of selected regions on the samples were measured using a SOC100 Hemispherical Directional Reflectometer (HDR) (Surface Optics Corporation) coupled with a FTIR Spectrometer at a sample temperature of \qty{20}{\degreeCelsius}. The incident polar angle is fixed at \ang{10}, near normal incidence. Kirchhoff’s law (\textalpha{} = \textepsilon) and assuming the transmissivity of an opaque material (\texttau{} = 0) enable the link between the hemispherical directional reflection factor \textrho{}  and the emissivity \textepsilon{}  for a given wavelength: 

\begin{equation}
    \varepsilon(\lambda)=\ 1-\rho(\lambda)\,.
\end{equation}

The total directional sample emissivity was then calculated from the ratio between the luminance L($\lambda$) radiated from the sample and the luminance L$^0(\lambda)$ from a black body integrated over the spectral range from \qtyrange{2.0}{30.2}{\micro\metre}.

\begin{equation}
    \varepsilon = \frac{ \int_{\lambda_1}^{\lambda_2} L(\lambda)\dd \lambda}{ \int_{\lambda_1}^{\lambda_2} L^0(\lambda)\dd \lambda }\,.
\end{equation}

\begin{table}[ht]
\small
    \centering
    \caption{Filter wheel configuration of the MS-IR camera.}
    \label{tab:camconfig}
    \begin{tabular}{lllrcclr}
\toprule
Position & Description & \multicolumn{2}{p{2cm}}{Spectral Range [\unit{\micro\metre}]} & OD & Transmittance [\unit{\percent}] & \multicolumn{2}{p{2cm}}{Temperature Range [\unit{\degreeCelsius}]} \\
\midrule
 FW\#1 & Sapphire window (Broadband) &     1.5 &      6.00 &        N/A &        100 &         10 &        338 \\
 FW\#2 & Neutral density filter, OD 1.6  &  1.5 &       6.00 &        1.6 &        2.2 &        207 &       1500 \\
 FW\#3 & Neutral density filter, OD 3 &     1.5 &       6.00 &          3.0 &        0.1 &        539 &       1500 \\
 FW\#4 & NPB-3010-40 (H2O) &     2.98 &      3.03 &        N/A &         70 &        408 &       1440 \\
 FW\#5 & NBP-4450-150 (CO2 and CO) &     4.31 &       4.61 &        N/A &         70 &        155 &        850 \\
 FW\#6 & NBP-3700-110 (Through flame) &  3.65 &       3.76 &        N/A &         70 &        225 &       1194 \\
 FW\#7 & NP-3800-040 (Through flame) &     3.78 &      3.82 &        N/A &         70 &        297 &       1500 \\
 FW\#8 & NB-3950-040 (Through flame) &     3.94 &       3.98 &        N/A &         60 &        282 &       1500 \\
\bottomrule
\end{tabular}  
\end{table}

\section{Theory and calculation}

The radiative power of a blackbody L$_\sigma\left(\sigma,T\right)$, an ideal material with an emissivity \textepsilon{} equal to unity at thermal equilibrium, depends on temperature and wavenumber as described by Planck’s law:

\begin{equation}\label{eq:planck}
    L_\sigma\left(\sigma,T\right)=2\cdot{10}^8\cdot hc^2 \sigma^3 \frac{1}{\exp{\left(\frac{100hc\sigma}{k_{\mathrm{B}}T}\right)}-1}  \, 
    \left[\unit{W/(m^2.sr.cm^{-1})}\right]
\end{equation}

With L$_\sigma\left(\sigma,T\right)$ the spectral radiance, \textsigma{} the wavenumber [\unit{cm^{-1}}], T the temperature [\unit{\kelvin}], h the Planck constant, c the speed of the light and k$_{\mathrm{B}}$ the Boltzmann constant. Integrating the spectral radiance over all wavenumbers yields the well-known Stefan-Boltzmann law, but for our application, only the spectral range of the camera for a given optical configuration is of interest. The in-band radiance (IBR) for a given filter from a cut-on, \textsigma$_{\mathrm{low}}$ to cut-off \textsigma$_{\mathrm{high}}$ wavenumber is mathematically linked to the black-body temperature by:

\begin{equation}\label{eq:ibrgeneral}
    \mathrm{IBR}(T)=\int_{\sigma_{\mathrm{low}}}^{\sigma_{\mathrm{high}}}{L_\sigma\left(\sigma,T\right)\dd \sigma}\, \left[\unit[per-mode = symbol]{\watt\per\square\metre\per\steradian}\right]
\end{equation}

A key feature of Eq.~\ref{eq:ibrgeneral} is that the function is strictly monotonic, thus allowing only one value of IBR per temperature for a given optical configuration. The radiometric camera calibration provides this unique link between an incoming radiance signal and a given black-body temperature. Through this relation, the radiometric temperature measured by a calibrated thermal infrared camera assumes an ideal black-body (BB) radiation L$_{\mathrm{BB}}(\sigma)$, from which real materials typically deviate, with the degree of deviation accounted for by their specific emissivity \textepsilon(\textsigma). This value must be obtained experimentally or from the literature to yield the corrected spectral radiance L(\textsigma) = L$_{\mathrm{BB}}$(\textsigma)\textepsilon(\textsigma). As the emissivity is a surface property, the important change in composition and morphology occurring as composite materials degrade has a direct effect on \textepsilon. For metallic samples, the formation of oxide layers and changes in surface finish may present an important source of error \cite{Hijazi2011, Jo2017}. In addition, spectral variations and temperature dependency make emissivity a material property that is difficult to assess, but easily changed by other material properties. 

Two-color (2C) pyrometry overcomes some of the challenges associated with obtaining accurate emissivity data. Instead of assuming an overall constant emissivity, it is only assumed constant between two narrow and closely located wavebands (\textsigma$_{1,  \mathrm{low}}$ to \textsigma$_{1,  \mathrm{high}}$, and \textsigma$_{2,  \mathrm{low}}$ to \textsigma$_{2,  \mathrm{high}}$). The IBR measured by the camera for a given spectral bands becomes:

\begin{equation}
    \mathrm{IBR}_{\sigma_i}(T)=\int_{\sigma_{i,\mathrm{low}}}^{\sigma_{i,\mathrm{high}}}{L_{\sigma_i}\left(\sigma_i,\mathrm{T}\right) \dd \sigma \, \left[\unit[per-mode = symbol]{\watt\per\square\metre\per\steradian}\right]\quad \mathrm{ with}\, i}\in\{1,2\}.
\end{equation}

Although the relation between temperature and IBR is nonlinear and may even vary depending on the spectral filter, it remains monotonic. Therefore, the ratio of the two IBR signals is monotonic as well. 2C pyrometry employs this mathematical property to eliminate the emissivity-dependance and to retrieve the true temperature of a material based on the IBR ratio. 

\subsection{Radiation model}\label{sec:radiation-model}

\begin{figure}[tb]
    \centering
    \includegraphics[width=0.55\textwidth]{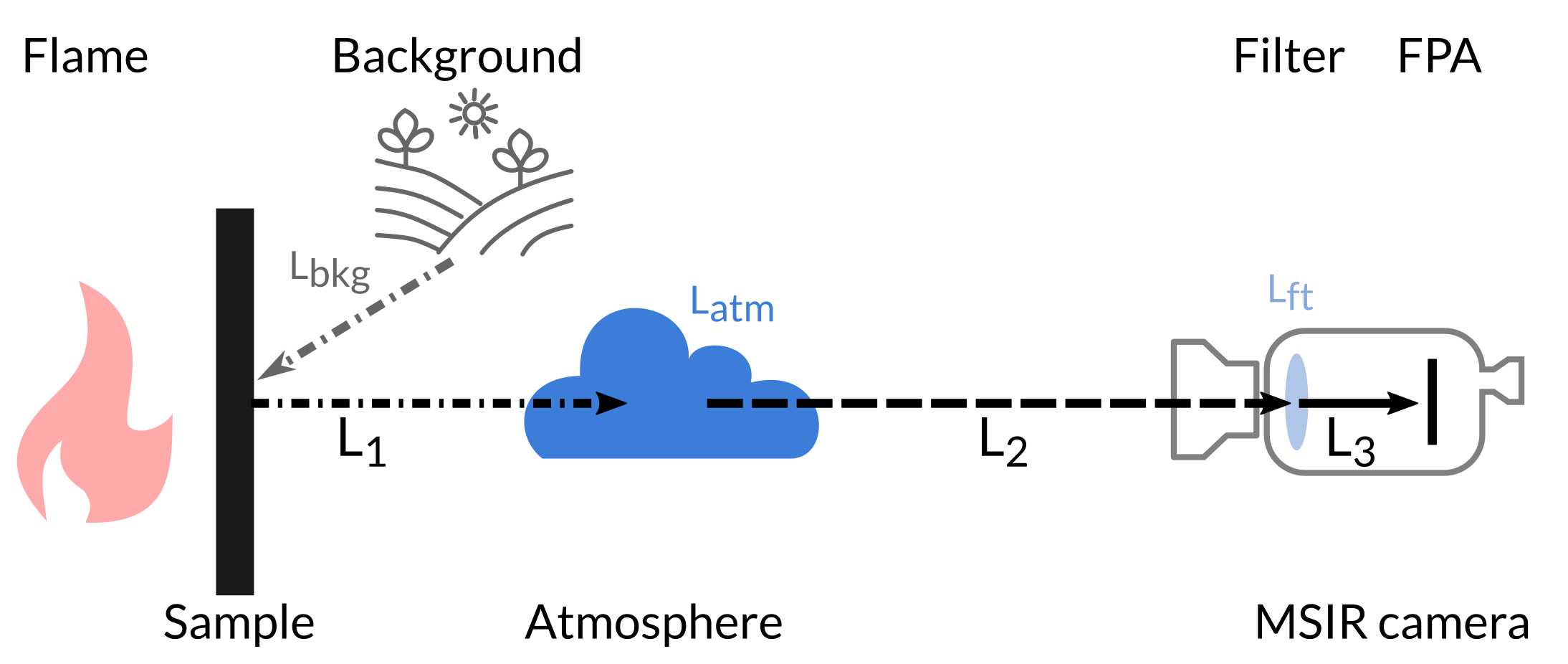}
    \caption{Schematic representation of the ray path  from sample to detector. The direct line of sight signal is affected by the filter characteristics and by potential contributions from the background and the atmosphere. The filter is installed in the internal wheel of the multi-spectra IR camera.}
    \label{fig:rmodel}
\end{figure}

The processing of IR image to extract temperatures requires proper assessment of the optical path (Fig.~\ref{fig:rmodel}) and of the factors influencing the total incident spectral radiance L$_{\sigma_i}(\sigma_i$,T) in Eq.~\ref{eq:planck} arriving at the detector in the different filter band for a given experimental setup. Phenomenologically, the optical path is described with a multi-layer radiative transfer model. The individual layers of the model for the configuration where a camera images the back (cold) face of a thin sample exposed to an open flame are depicted in Fig.~\ref{fig:rmodel}. It includes contributions from the background (bkg) reflected onto the sample (s), the pilot flame (fl) as well as the atmospheric (atm) and filter (ft) transmission characteristics. The radiative properties of each layer are derived according to the first law of thermodynamics, where absorptivity, reflectance, and transmittance of a material sum to unity, \textalpha+\textrho+\texttau=1. Furthermore, assuming diffusive gray body behaviour, the relation can be rewritten based on Kirchhoff’s law, as the absorptivity directly equals the emissivity, \textalpha$_s$ = \textepsilon$_s$. The radiative properties of the contributing layers are summarized in Tab.~\ref{tab:rmodel}.

\begin{table}[hb]
\centering
\caption{Radiative contributions from the model components shown in Fig.~\ref{fig:rmodel}.}\label{tab:rmodel}
\begin{tabular}{llll}
\toprule
Model component & negligible quantity & energy balance & contribution\\
\midrule
Background & \texttau{} = 0 & \textalpha{} = 1 & \textepsilon$_{\mathrm{bkg}}$ = 1 \\
Atmosphere & \textrho{} = 0 & \texttau{}+ \textalpha{} = 1 & \textepsilon$_{\mathrm{atm}}$ = 1 - $\tau_{\mathrm{atm}}$\\
Filter &     \textrho{} = 0 & \texttau{}+ \textalpha{} = 1  & \textepsilon$_{\mathrm{ft}}$ = 1 - $\tau_{\mathrm{ft}}$\\
\bottomrule
\end{tabular}
\end{table}

The radiance L$_1$(\textsigma,T) close to the sample surface is determined by the sample spectral emissivity \textepsilon$_{\text{s}}$(\textsigma) and temperature T$_{\text{s}}$ as well as by potential background contributions L$_{\text{bkg}}$ reflected to the sample:

\begin{equation}
    L_1\left(\sigma,T\right) = \varepsilon_{\text{s}}\left(\sigma\right) L_{\text{s}}\left(\sigma,T_{\text{s}}\right)+\left(1-\varepsilon_{\text{s}}\left(\sigma\right)\right){L}_{\text{bkg}}\left(\sigma,T_{\text{bkg}}\right)\, \left[\unit[per-mode = symbol]{\watt\per(m^2.sr.{cm}^{-1})}\right] 
\end{equation}

The atmosphere can alter the total radiance received at the detector, especially if species with strong absorbance bands in the IR spectrum are present, such as water vapour and carbon dioxide:

\begin{equation}
    L_2\left(\sigma,T\right)=\left(1-\tau_{\text{atm}}\left(\sigma\right)\right) L_{\text{atm}}\left(\sigma,T_{\text{atm}}\right)+\tau_{\text{atm}}\left(\sigma\right) L_1\left(\sigma,T\right)\ \left[\unit{W/(m^2.sr.{cm}^{-1})}\right]
\end{equation}

The camera band-pass filters characteristics finally determine the total spectral radiance incident on the camera detector: 

\begin{equation}\label{eq:L3}
    L_3\left(\sigma,T\right)=\left(1-\tau_{\text{ft}}\left(\sigma\right)\right) L_{\text{ft}}\left(\sigma,T_{\text{ft}}\right)+\tau_{\text{ft}}\left(\sigma\right) L_2\left(\sigma,T\right)\ \left[\unit{\watt/(m^2.sr.{cm}^{-1}})\right]
\end{equation}

In the case of an external filter wheel and un-cooled camera used in other setups \cite{Reggeti_Agrawal_Bittle_2019, Sanchez-Carballido2017, Melendez2010, Arakawa_Saito_Gruver_1993, Ballester2010}, additional contributions from the environment reflected by the filter outside of its transmissivity range would have to be considered. In our experiments, the filters are installed in an internal wheel located within the cooled MS-IR camera (Fig.~\ref{fig:msir}), which renders the radiation reflected negligible relative to the direct line-of-sight signal \cite{Savino2017}. Eq.~\ref{eq:L3} contains seven unknowns and integrating over the spectral range of the camera adds two more parameters, while only a single value, the total integrated band radiance of the camera IBR$_{\text{cam}}$ is accessible. Radiometric calibration, filter specifications and well-justified assumptions allow the model to be reduced to two unknowns, the sample temperature and emissivity. First, the radiometric camera calibration from the supplier already considers the band-pass filter response (Fig.~\ref{fig:filter}) in the spectral and temperature region of interest and provides a direct access to the incoming radiance to the camera:

\begin{equation}\label{eq:IBRcam}
    \text{IBR}_{\text{cam}}(T) = \int_{\sigma_{\text{low}}}^{\sigma_{\text{high}}}{L_3\left(\sigma,T\right) d\sigma}\ \left[\unit[per-mode=symbol]{\watt\per\square\metre\per\steradian}\right]
\end{equation}

Secondly, band-pass filters are chosen to be transparent to the atmosphere at short distances (<\qty{1}{\metre}) from the target and the camera is place at \qty{85}{\centi\metre} from the sample surface. It is thus assumed that both, \texttau$_{\text{atm}}$(\textsigma) and \texttau$_{\text{ft}}$(\textsigma), are equal to unity over the full spectral range. With this, L$_3$(\textsigma,T) = L$_2$(\textsigma,T), and L$_2$(\textsigma,T) = L$_1$(\textsigma,T), and Eq.~\ref{eq:IBRcam} becomes:

\begin{equation}
    \label{eq:IBRcam2}
    \text{IBR}_{\text{cam}}(T)=\int_{\sigma_{\text{low}}}^{\sigma_{\text{high}}}{L_1\left(\sigma,T\right)\dd \sigma}\ \left[\unit[per-mode=symbol]{\watt\per\square\metre\per\steradian}\right]
\end{equation}

With a band gap as small as \qty{0.04}{\micro\metre} for FW\#7 and \qty{0.08}{\micro\metre} for FW\#8, emissivity is assumed constant over these two spectral ranges. With this important simplification, \textepsilon$_{\text{s}}$(\textsigma) = \textepsilon$_{\text{s}}$ and Eq.~\ref{eq:IBRcam2} now becomes: 

\begin{equation}
    \label{eq:IBRcam3}
    \text{IBR}_{\text{cam}}(T)=\int_{\sigma_{\text{low}}}^{\sigma_{\text{high}}}{\left[\varepsilon_s L_s\left(\sigma,T_s\right)+\left(1-\varepsilon_s\right){L}_{\text{bkg}}\left(\sigma,T_{\text{bkg}}\right)\right]\dd \sigma}\ \left[\unit[per-mode=symbol]{\watt\per\square\metre\per\steradian}\right]
\end{equation}

The surface emissivity \textepsilon$_s$ also determines the relative importance of the reflected background radiance, but this contribution may be disregarded if the sample temperature is significantly higher than its background, which is generally the case in the fire certification test considered here. As a rule of thumb, the background radiance should be less than one percent of the sample radiance \cite{Vollmer_Mollmann_2018}. As a first estimate, we assumed an average background temperature of \qty{300}{\kelvin}, an emissivity of \textepsilon{}=~0.6 for metal samples, and \textepsilon{}=~0.8 for composite samples. Aiming for a signal-to-noise ratio of 100, for the filters with a centre wavelength \textlambda{} =~\qty{3.7}{\micro\metre}, the minimum sample temperature above which the background contribution can be disregarded is approximately \qty{440}{\kelvin} (\qty{167}{\degreeCelsius}) for metals and \qty{400}{\kelvin} (\qty{127}{\degreeCelsius}) for composite samples. Both temperature values lie well below the minimum temperature of the filters FW\#7 and FW\#8 used for our tests (cf. to Tab.~\ref{tab:camconfig}). With these assumptions the transfer model from Eq.~\ref{eq:IBRcam3} can be further simplified:

\begin{equation}\label{eq:IBRcamfinal}
    \text{IBR}_{\text{cam}}(T)=\varepsilon_s\times\int_{\sigma_{\text{low}}}^{\sigma_{\text{high}}}{L_s\left(\sigma,T_s\right) \dd \sigma} \left[\unit[per-mode=symbol]{\watt\per\square\metre\per\steradian}\right]
\end{equation}

Finally, the assumption that led to Eq.~\ref{eq:IBRcam3} is generalized. Since the two filters bands are close to each other by design, the sample emissivity is not only considered invariable in the spectral range of each filter, but also assumed to be the same for both filters. The initial radiation model with nine unknowns can thus be reduced to a system of only two unknowns, IBR$_{\text{filter, 1}}$ and IBR$_{\text{filter, 2}}$, which are both experimentally accessible. Applying the 2C pyrometry principle, we use Eq.~\ref{eq:IBRcamfinal} to form the IBR signal ratio for two filters, an expression independent of the sample emissivity:

\begin{equation}\label{eq:ibr-ratio}
    r\left(T_s\right)=\frac{\text{IBR}_{\text{filter1}}}{\text{IBR}_{\text{filter2}}}\, \left[-\right]
\end{equation}

\begin{figure}
    \centering
    \includegraphics[width=0.45\textwidth]{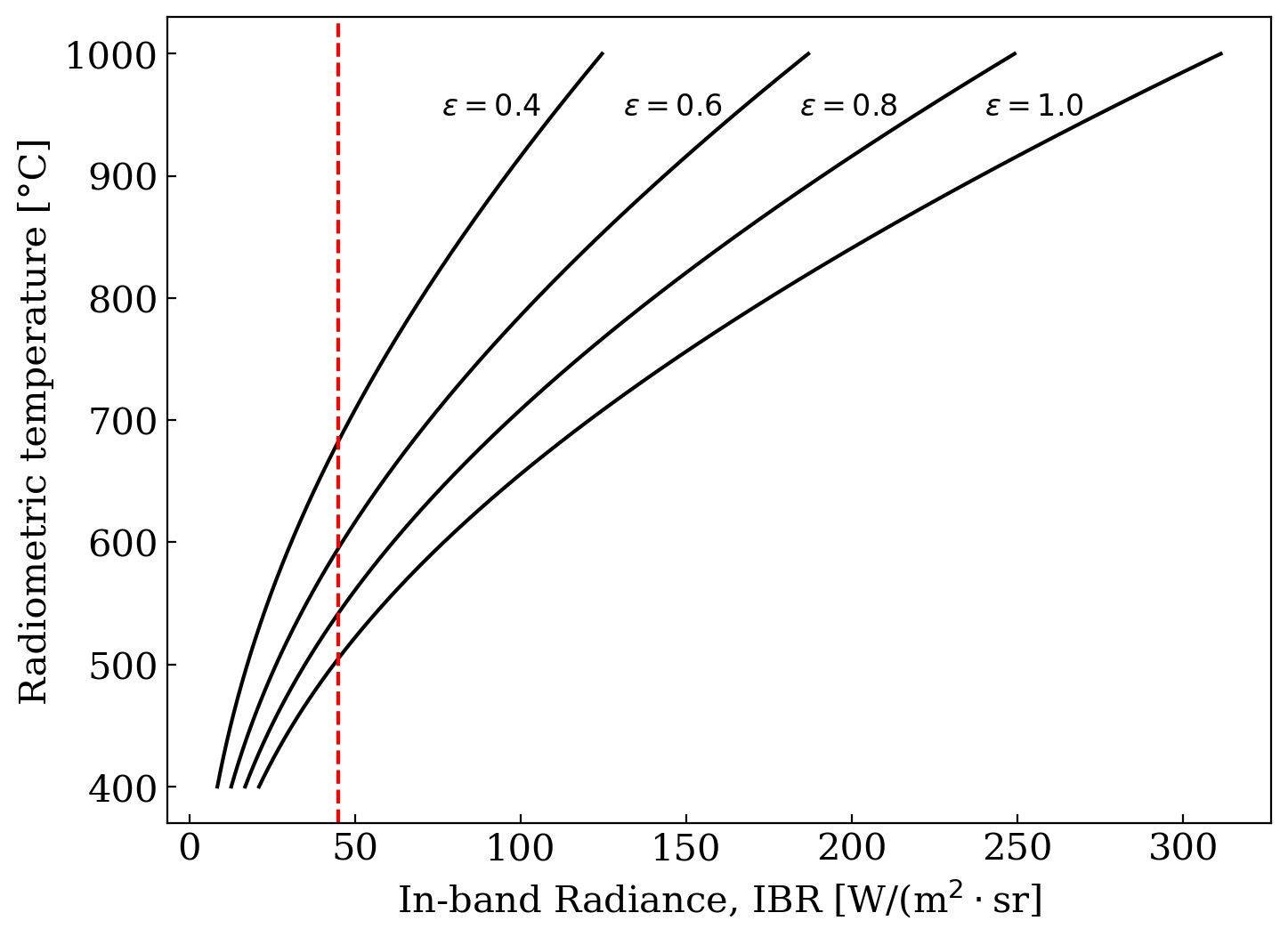}
    \caption{Simulated temperature curves for filter FW\#7 (cf. to Tab.~\ref{tab:camconfig}) as a function of IBR for different sample emissivities \textepsilon{} demonstrating the monotonous relationship between. The red line highlights was added as an example to illustrate that a wrong emissivity value results in wrong temperature readings.}
    \label{fig:ibremiss}
\end{figure}

The monotonicity of Eq.~\ref{eq:IBRcamfinal} ensures that Eq.~\ref{eq:ibr-ratio} is also monotonic and hence that any given temperature for a sample with emissivity \textepsilon$_s$ corresponds to a distinct IBR ratio value. Simulated IBR-temperature curves for FW\#7 with different \textepsilon{} are shown in Fig.~\ref{fig:ibremiss}. Using an exemplary value of \qty[per-mode=symbol]{50}{\kW\per\square\m\per\steradian} (red line), it becomes immediately apparent that errors in emissivity can easily result in wrong temperature readings when relying on single colour (1C) pyrometry. Wrongly assuming an emissivity of 0.8 for a sample with \textepsilon{}= 0.6, the sample temperature is underestimated by \qty{53}{\degreeCelsius} in our example. 

The practical implementation of the 2C techniques is referred to hereafter as the Temperature-Emissivity-Separation (TES) algorithm and relies on several steps.
First, the IBR ratio in Eq.~\ref{eq:ibr-ratio} for two different filters is simulated by numerically solving Eq.~\ref{eq:IBRcamfinal} over the respective filter bands and over a temperature range of interest. Then, the IBR ratio obtained from the actual camera readings IBR$_{\text{filter1}}$ and IBR$_{\text{filter2}}$ is compared with the simulation to identify the corresponding sample temperature. In a last step, this sample temperature can be used to calculate the corresponding spectral radiance for a black body based on Eq.~\ref{eq:IBRcamfinal} and using the information from a single filter channel. This allows to calculate the sample emissivity:  

\begin{equation}
    \label{eq:eps-definition}
    \varepsilon_{s,i}=\frac{\text{IBR}_{\text{filter,i}}}{\int_{\sigma_{i,\,\text{low}}}^{\sigma_{i,\,\text{high}}} {L_s\left(\sigma,T_s\right) \dd \sigma}}\, [-]\,.
\end{equation}

Because the IBR ratio, r(T$_s$) is built pixel-wise, the TES algorithm allows to obtain sample temperature and emissivity mappings with the size and resolution determined by the exposed detector area and distance to the sample. 

\subsection{Sources of error}
The principle of 2C pyrometry allows to separate temperature and emissivity calculations for the same scene using two images acquired with different band-pass filters.  The apparent simplicity of this technique to obtain both quantities from the same measurement comes with several key requirements for the acquired IR signal that are implied by the choice of radiation model and assumptions described in Sec.~\ref{sec:radiation-model}. Any deviation from the ideal case contributes to the overall error budget.

Starting right-to-left with respect to Fig.~\ref{fig:rmodel}, three basic limitations are linked to the camera hardware: mechanical precision, attainable speed of the fast-rotating filter wheel and minimal detector noise. The mechanical transition from one filter to another induces a parallax error because of the bore-sight displacement. A fine co-registration is crucial for pixel-based IBR calculations, otherwise small features are more prone to accumulate errors at low spatial resolution and a dedicated post-processing step is recommended. For the metal and PMC samples tested in this study, the focus was set on the entire sample region visible or on a larger sub-region so that the resolution was never below \qty{4}{\milli\meter\per\pixel}. The filter wheel mechanics are responsible for filter positioning and image alignment, while ensuring stable and fast rotation. 
A critical potential source of error in fire testing, where temperature and material properties can change quickly, is temporal resolution. In this situation, one is typically interested in following rapid changes in temperature as, for example, during the first few tens of seconds when a PCM is heated and ignited. To accurately track the temperature, the delay caused by the filter wheel turning needs to be small enough for the camera IBR signal to be considered invariant between subsequent acquisitions. To this end, the camera acquisition frame rate is synchronized with the filter wheel rotation and the filters of interest are mounted in adjacent positions. The IBR-ratio is calculated for the successive frames for each filter wheel rotation where any signal variation contributes to the overall error budget. The selection of an appropriate acquisition frequency is thus an essential part of the experimental design. The frame rate is limited by a combination of window size, exposure time, and the fastest (mechanically) attainable wheel speed.

The temporal noise at the detector impacts the pixel-wise error for each frame and shall be limited by the camera shot noise. For this requirement, the camera manufacturer usually refers to the noise equivalent temperature difference (NETD) at room temperature. The NETD can be expressed in terms of radiance with the noise equivalent differential in-band radiance (NEdIBR). This formulation provides a useful performance figure for each filter within a specific temperature range as to the amount of change in radiance needed to equal the system noise. High-quality band-pass filters with a narrow gap and sharp cut-on and cut-off characteristics increase the radiometric accuracy. Neglecting test specific error sources, the ultimate performance limit of 2C pyrometry is therefore governed by the camera hardware (detector, filter, rotating wheel) and the resulting camera specific NEdIBR.

\begin{figure}[tb]
    \centering
    \begin{subfigure}[t]{0.45\textwidth}
    \centering
         \includegraphics[width=\textwidth]{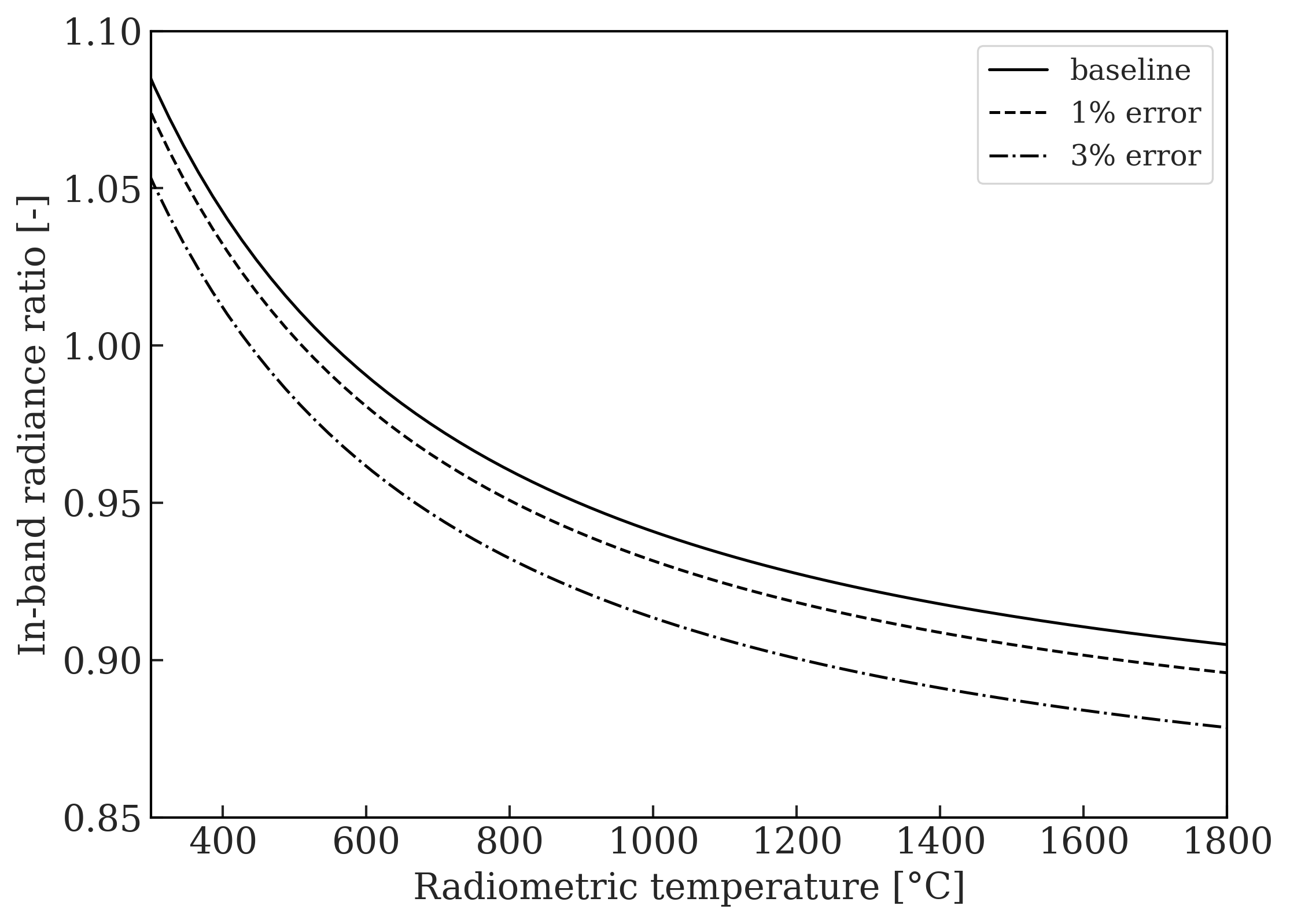}
    \caption{}
    \label{fig:IBR-RT}
    \end{subfigure}
    \begin{subfigure}[t]{0.45\textwidth}
    \centering
         \includegraphics[width=\textwidth]{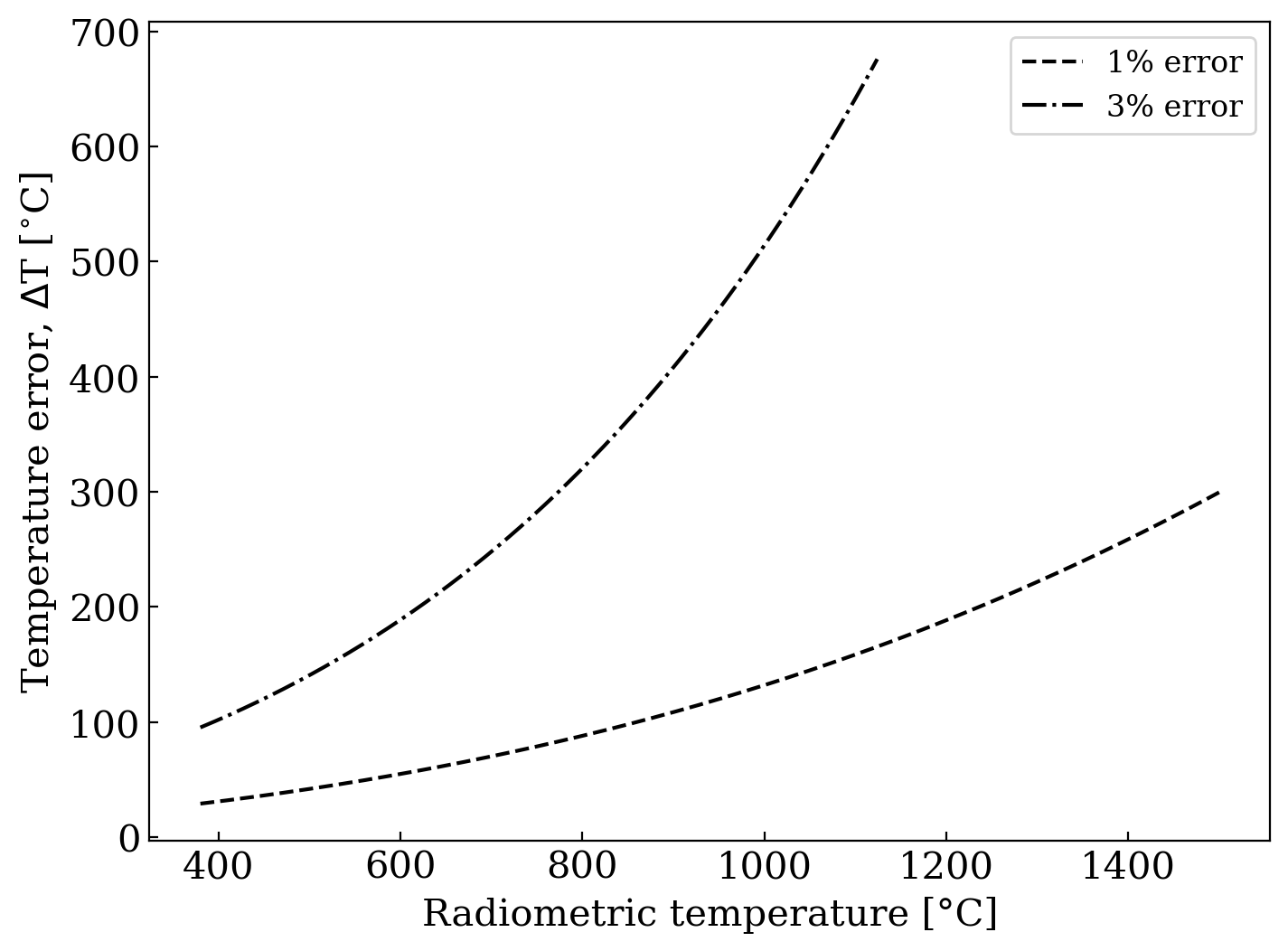}
    \caption{}
    \label{fig:dT-RT}
    \end{subfigure}
        \caption{(a) Simulated in-band radiance ratio (IBR) signal for filters FW\#7 and FW\#8 over the temperature range of interest. A deviation of \qty{1}{\percent} and \qty{3}{\percent} from the baseline signal is shown. (b) Assuming an otherwise perfect measurement system, these deviations result in large errors especially at high temperatures.}
    \label{fig:error-effects}
\end{figure}

The challenge to quantify both temperature and emissivity becomes immediately apparent when considering a small error in one of the spectral bands. This error can be caused by a calibration inaccuracy for a filter, by variations of the radiance signal within a filter’s spectral band or in-between acquisitions. Using the spectral bands for filters FW\#7 and FW\#8 used in our experiment (cf. to Tab.~\ref{tab:camconfig}), the calculated temperature variation due to an error of \qty{1}{\percent} or \qty{3}{\percent} in the spectral band of filter FW\#7 is simulated in Fig.~\ref{fig:error-effects}. At \qty{500}{\degreeCelsius}, a common upper limit for backside temperatures of PCM fire testing, the temperature is underestimated by approximately \qty{43}{\degreeCelsius} and more than \qty{140}{\degreeCelsius}, for \qty{1}{\percent} and \qty{3}{\percent} filter errors, respectively. For PMCs, the temperature regime around \qty{500}{\degreeCelsius} is critical, as many important processes occur, such as resin decomposition, degradation of the fibre sizing agent and fibre oxidation \cite{Gibson_Torres_Browne_Feih_Mouritz_2010, Swanson2018}. Aiming for a measurement accuracy of \qty{10}{\degreeCelsius} for each contributor to the overall error budget requires the radiance variation between two filters to be as low as \qty{0.25}{\percent}. The same requirement applies to the radiometric accuracy of the camera calibration. 

\begin{figure}
    \centering
    \includegraphics[width=0.75\textwidth]{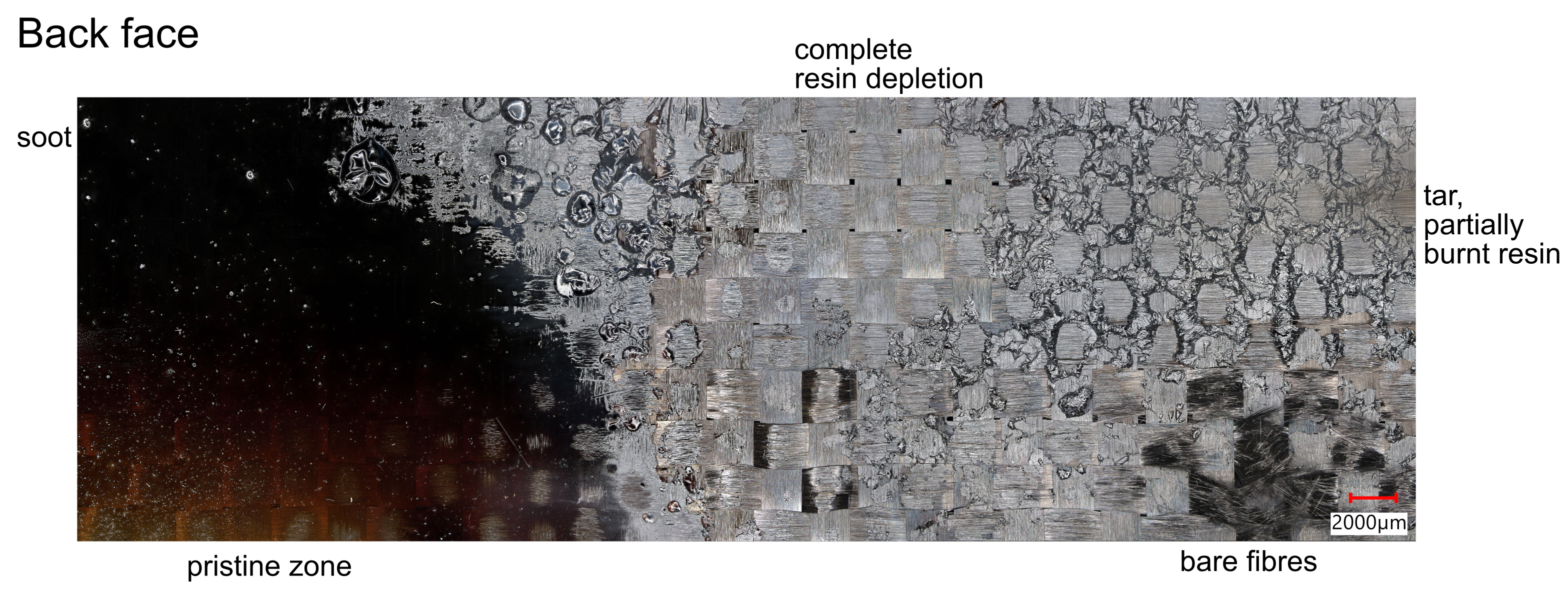}
    \caption{Back face (BF) of a carbon fiber epoxy sample strips after a \qty{30}{\second} fire test using a kerosene flame, revealing highly variable surface properties. The sample strip shown is \qtyproduct{25.4 x 80.0}{\milli\metre}.}
    \label{fig:burntsampleregion}
\end{figure}

Finally, the samples are required to behave as gray bodies with an emissivity that is independent of wavelength, temperature, and solid angle in both filter bands, \textepsilon$_s$(\textlambda, T, \textTheta) = \textepsilon$_s$. The crux of the problem for PMCs is that the “body” thermally degrades into several different materials, as shown in Fig.~\ref{fig:burntsampleregion} from a partially burnt carbon-fibre epoxy sample exposed for \qty{30}{\second} to a kerosene flame calibrated to \qty[per-mode=symbol]{116}{\kilo\watt\per\square\metre}, representative for aerospace large scale fire tests \cite{ac20135,iso2685}. Already after this short flame exposure, distinct zones with likely different emissivities can be observed on the back side as the resin partially burns and decomposes. For steel, the surface undergoes an important emissivity change due to oxidation but once the oxide layer is established the temperature dependence of the target emissivity is sufficiently subtle and can usually be neglected \cite{Saunders_2001}. 

In general, the total emissivity for many materials shows a strong temperature dependency. Additionally, the change of emissivity with temperature is often correlated to its change with wavelength. A surface temperature rise for a sample means also that its emission is shifted to shorter wavelengths. In general, \textepsilon(T) increases with T for conductors and decreases for insulators \cite{Jones_Mason_Williams_2019}. Although electrically conductive, an emissivity decrease with temperature has also been reported for carbon fibre composites \cite{Li_Strieder_2009, Zhang_Dai_Lu_Wu_2016, Athanasopoulos2012}. Some authors \cite{Li_Strieder_2009, Zhang_Dai_Lu_Wu_2016, Jones_Mason_Williams_2019} propose a second-order polynomial fit for the \textepsilon(T) relation, with the coefficients based on literature data that remains valid only within a strict temperature range. Although such a fit certainly does not provide much insight into the physics at play it allows for a simple judgment method to compare the  data presented in Sec.~\ref{sec:insitu-em} against similar experimental studies that attempt to take into account the temperature dependence of emissivity.

\section{Image postprocessing}

\begin{figure}[ht]
    \centering
    \includegraphics[width=0.95\textwidth]{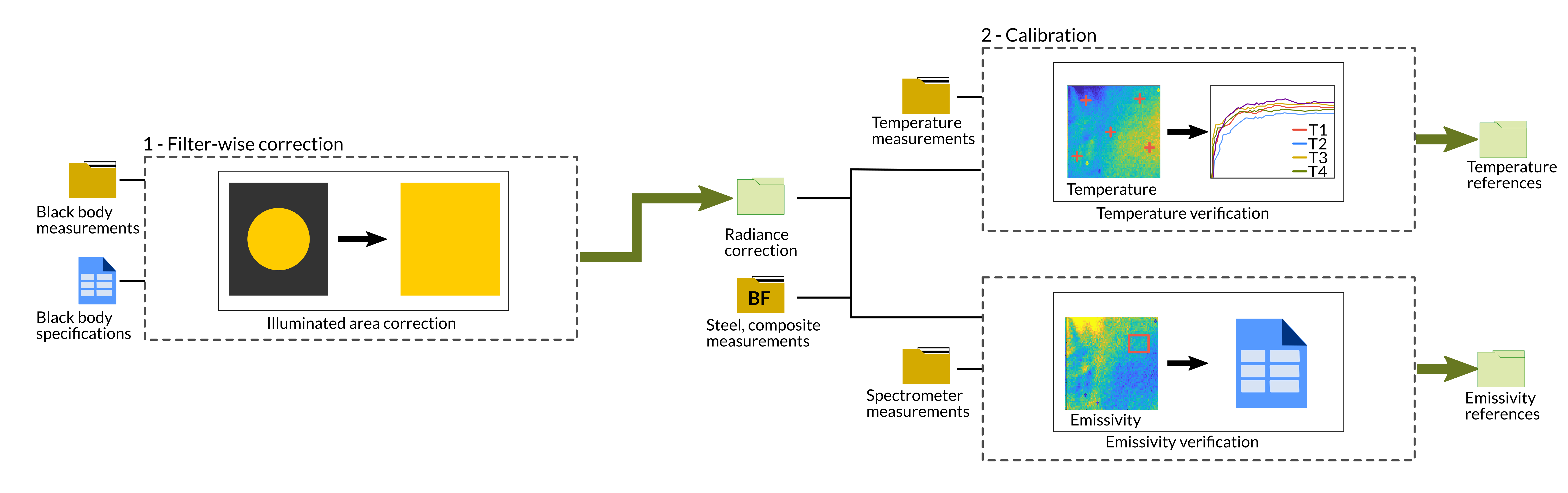} 
    \caption{Diagram of the different steps of the 2C thermography post-processing technique. Each section is linked to a dedicated experimental design and test campaign.}
    \label{fig:imageproc}
\end{figure}

The TES algorithm uses the IR raw signals from filter channel FW\#7 and FW\#8 to calculate a pixel-wise IBR ratio. This ratio is then used to retrieve the corresponding radiometric temperature from look-up tables. The post-processing accounts for the simplifications that have led from Eq.~\ref{eq:L3} to Eq.~\ref{eq:IBRcamfinal} by dedicated measurement blocks.  
The MS-IR camera records the IBR signal per pixel and registers the sequences in a proprietary Telops format (.hcc). Signal post-processing was performed using Matlab{\small\textregistered} 2019 and the Python programming environment (Python Software Foundation). The analysis workflow is depicted in Fig.~\ref{fig:imageproc} with each independent measurement block being grouped by a dashed line: radiance correction, temperature calibration, and emissivity evaluation. The filter-wise radiance correction are important, but independent steps when measuring thermally degrading complex materials. The calibration step allows a “proof of concept” and helped us to compare the TES algorithm results against independently measured values, relying on thermocouples.   

Part of the filter-wise correction account for the radiance of the detector area actually illuminated. This step is necessary as the camera calibration by the manufacturer is based on black body measurements with all sensor pixels exposed to a similar radiant flux. For tests with composite coupons strips, the camera field of view (FOV) was chosen to cover the vertical sample length that corresponds to the diameter of the stagnation zone of the impinging flame. The numerical simulations based on Eq.\ref{eq:IBRcamfinal} rely on the manufacturer calibration of the camera for each filter which is based on a homogeneously illuminating each pixel of the sensor.  Black-body measurements are used to create a look-up table that relates the number of illuminated pixels to a suitable radiance correction factor. It is important to note that each filter needs to be corrected independently. The calibration block tests, the gray body assumption as well as the temperature and emissivity mapping are compared against HDR and TC measurements. Since the HDR emissivity measurements were performed on \qtyproduct{25.4 x 25.4}{\milli\metre} cut-out samples after fire exposure, the TES algorithm used only the respective regions of interest (ROI) of the same samples on a subset of frames at the end of acquisition. The emissivity thus calculated was then averaged over the sub-region and frame set. In a similar manner, temperature mapping in a small pixel region at the estimated TC position was extracted over all frames and compared against the TC data. 

\section{Results and discussion}

\subsection{Filter-wise radiometric correction}\label{sec:rc}

\begin{figure}[tb]
    \centering
    \begin{subfigure}[t]{0.5\textwidth}
    \centering
        \includegraphics[width=0.8\textwidth]{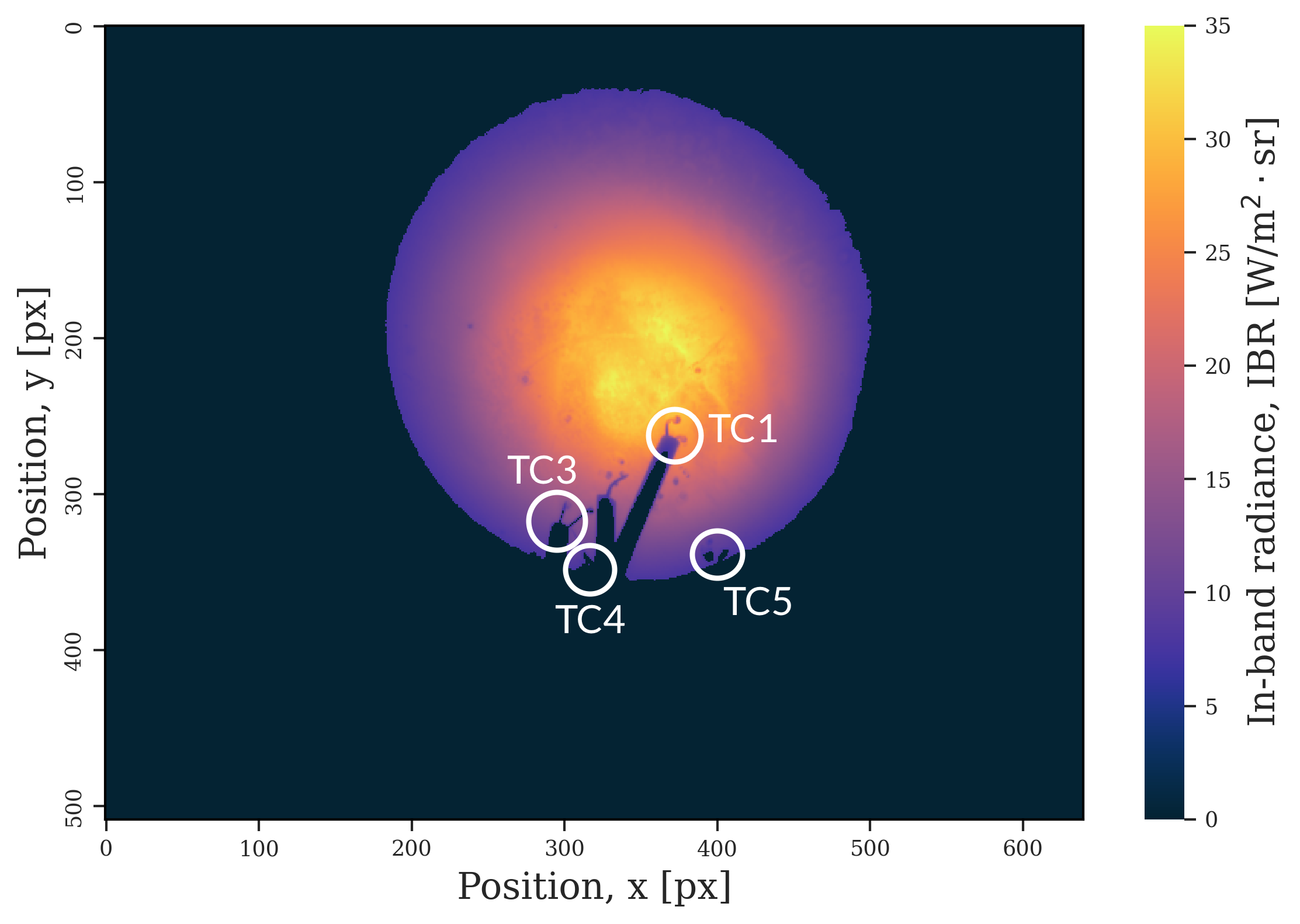}
        \caption{}\label{fig:steelIR}
    \end{subfigure}
    \begin{subfigure}[t]{0.45\textwidth}
    \centering
        \includegraphics[width=0.9\textwidth]{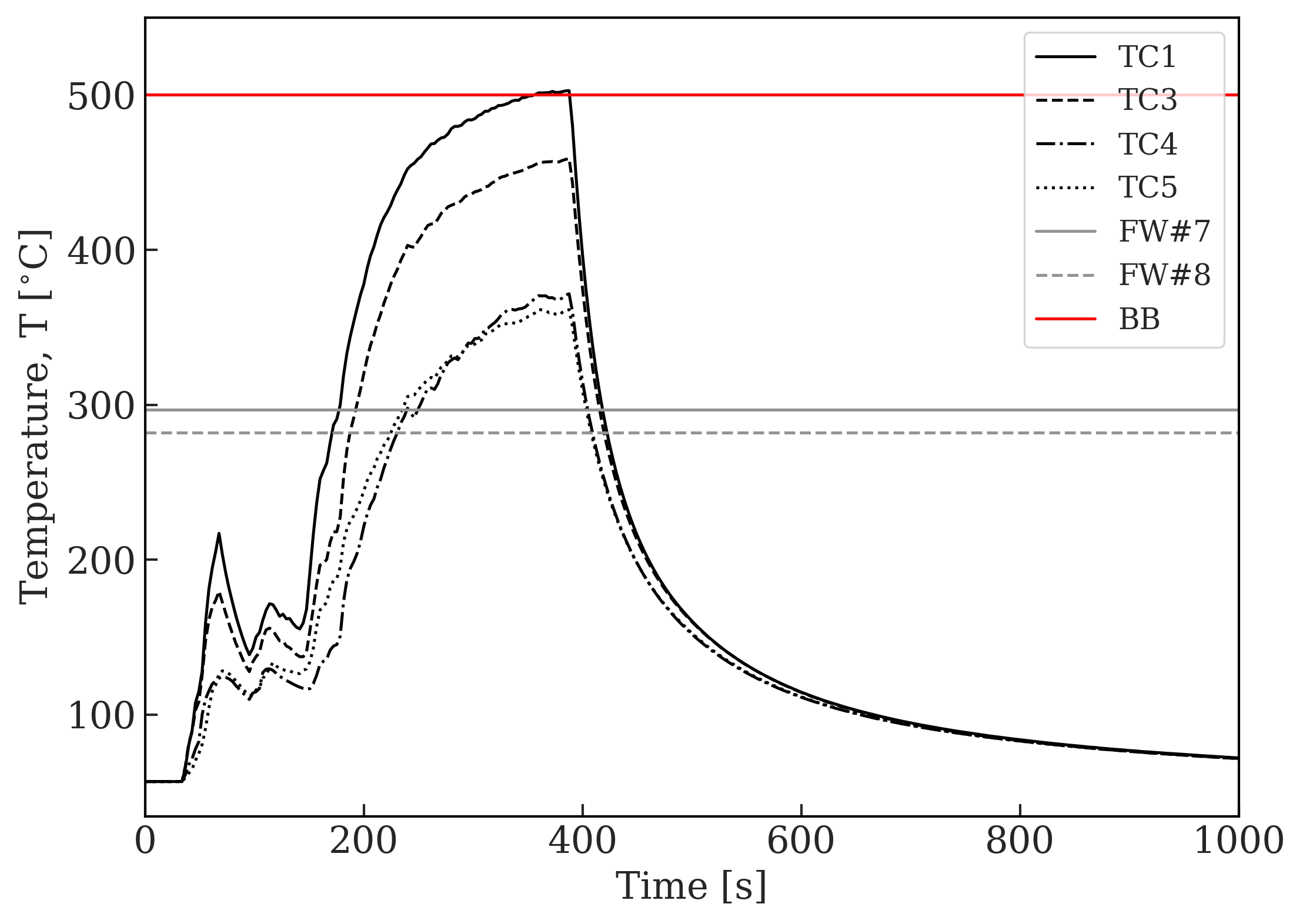}
        \caption{}\label{fig:steelTC}
    \end{subfigure}
    \caption{IR image of the backside of the steel plate, heated with a gas torch from the opposite site, with spot-welded thermocouples and their ceramic tubes partly visible. (b) Thermocouple temperature curves during a test, reaching approximately \qty{500}{\degreeCelsius} in the centre. TC4 and TC5 are located at the same radial distance (\qty{30}{\mm}) from the centre. The thermocouple accuracy is \qty{\pm0.75}{\percent} in this range and does not exceed \qty{\pm 3}{\degreeCelsius}. The lower threshold temperatures for FW\#7 and FW\#8 are indicated with horizontal lines.}
    \label{fig:steelplate}
\end{figure}

In a first step, a gas flame (2.7~slpm C$_3$H$_8$ and 8~slpm oxidizer mixture consisting of \qty{60}{\percent} O$_2$, diluted with CO$_2$) was used to heat a \qtyproduct{305 x 305}{\milli\metre}  stainless-steel sample with thermocouples welded at different radial distances from the stagnation point on its backside. A rich flame (\textPhi=1.3) is used to create a homogeneously heated zone with a temperature maximum below or equal to \qty{500}{\degreeCelsius} that can be recreated with the black-body. The IR image of the steel back face and the temperature curves measured by the TC in the FOV are shown in Fig.~\ref{fig:steelplate}. The pixels above the filter specific radiance threshold (Fig.~\ref{fig:steelIR}) corresponds approximately to a circular zone with a diameter of \qty{75}{\milli\metre} and an area of \qty{35000}{\square\pixel}. Fig.~\ref{fig:steelTC} shows the corresponding minimum temperatures for both filters as horizontal lines. Similar tests were repeated with flames of different diameter.

\begin{figure}[ht]
    \centering
        \includegraphics[width=0.45\textwidth]{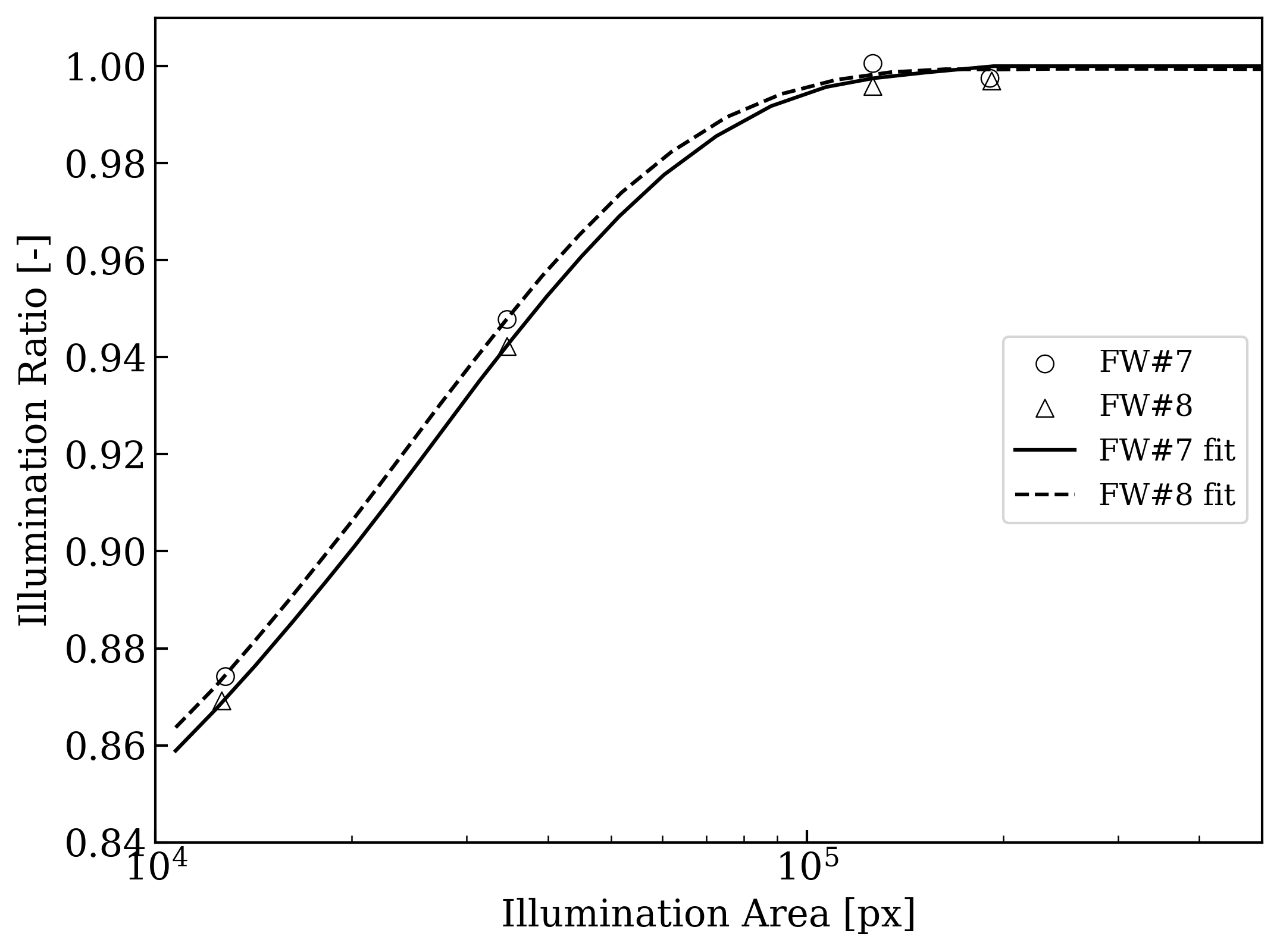}
    \caption{Radiance correction based on the contributing pixels with respect to a fully illuminated sensor array.}\label{fig:rc}
\end{figure}

The calibration starts by comparing the internal camera calibration against a black-body (set at \qty{500}{\degreeCelsius}) with a similar apparent temperature and FOV as the exposed sample during the experiment. To this end, the black-body was partially shielded, with the size of the illuminated area (IA) increased from \qtyrange{4}{58}{\percent}. The IA represents the fraction of total pixels \qtyproduct{640x512}{\pixel} contributing to the total signal. The results are listed in Tab.~\ref{tab:radcorr}.  Radiance correction (rc) is the inverse of the radiance ratio between the full aperture and a given IA. 
The IA-radiance relation is highly nonlinear, as can be seen in Fig.~\ref{fig:rc}. For instance, an IA of \qty{10}{\percent} yields an incoming signal through for filter FW\#7 that is only \qty{91}{\percent} of the expected radiance signal for a body of the same temperature and emissivity. As a result, without correction temperatures will be underestimated and emissivity will be overestimated, and hence a geometrical correction is essential for accurate 2C pyrometry measurements over a surface with steep temperature gradients.

\begin{table}[hb]
\centering
\caption{Radiance correction for the two filters, FW\#7 and FW\#8, obtained from black body measurements. The correction values (rc-FW\#7, rc-FW\#8) are filter-wise radiance correction for the illuminated area in the field of view.}\label{tab:radcorr}
\begin{tabular}{lccccc}
\toprule
IA [\unit{\percent}]& IA [\unit{\pixel}] &\textepsilon{}[-] & T [\unit{\degreeCelsius}] & rc-FW\#7 & rc-FW\#8 \\
\midrule
4 &		12679   & 0.993 &	492.0 &	1.151 &	1.144  \\
10 &	34419	& 0.910 &	500.6 &	1.061 &	1.055  \\
38 &	125195	& 0.986 &	489.1 &	1.003 &	1.000  \\
58 &	190468	& 0.988 &	492.3 &	1.003 &	1.002  \\
\bottomrule
\end{tabular}
\end{table}

\subsection{Post-fire emissivity measurements}\label{sec:emHDR}

\begin{figure}[tb]
    \centering
    \includegraphics[width=.7\textwidth]{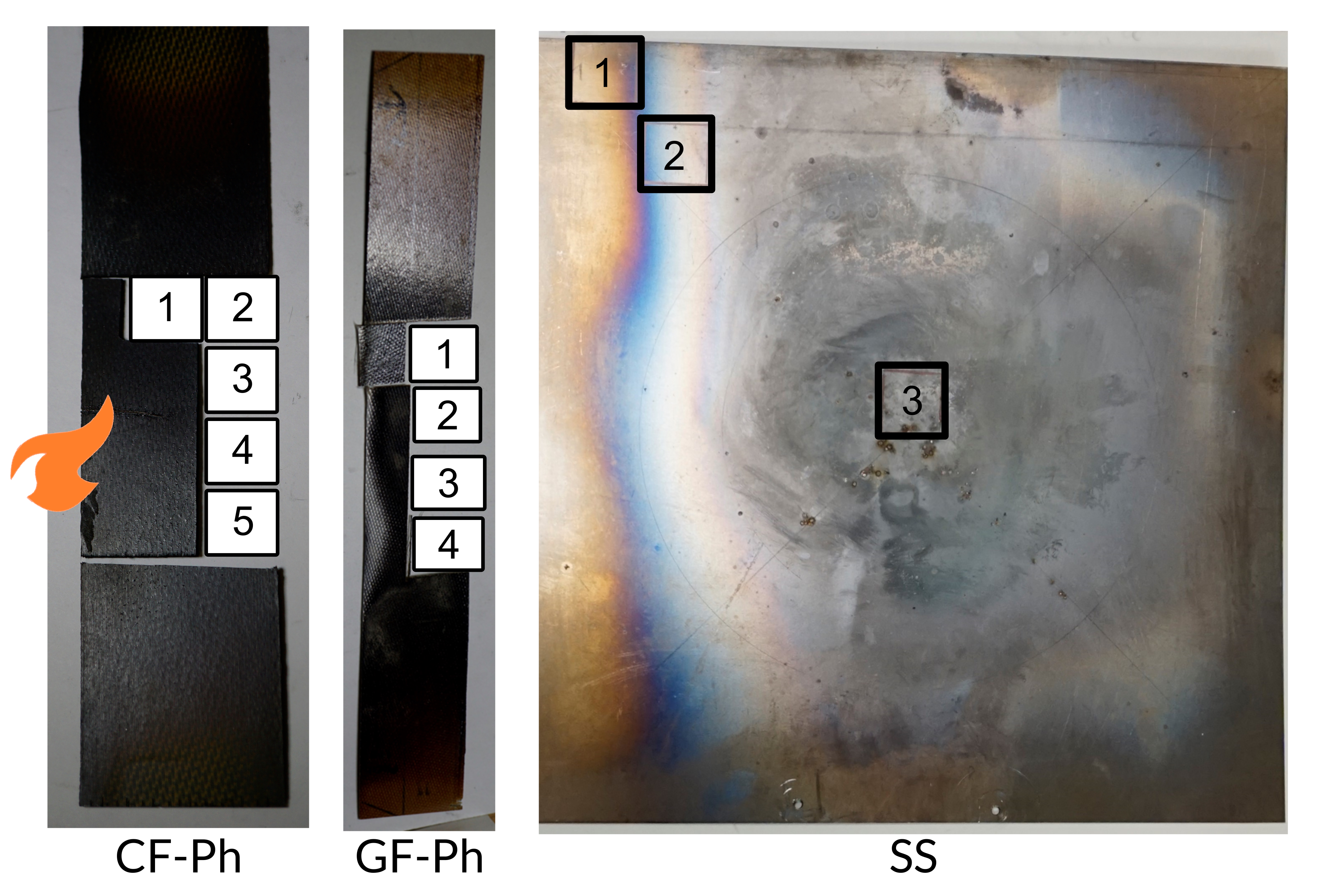}
    \caption{Samples used for post-fire emissivity measurements. The flame impingement was centred at the vertically oriented samples in-between position two and three for the GF-Ph and in-between position three and four for the CF-Ph sample. \qtyproduct{25.4 x 25.4}{\milli\metre} squares have been cut from the samples for emissivity measurements.}
    \label{fig:samplecuts}
\end{figure}

The insitu emissivity values obtained with the TES algorithm are compared against an independent measurement technique implemented at room temperature following fire exposure. A kerosene flame (calibrated to \qty[per-mode=symbol]{116}{\kilo\watt\per\square\metre}) was used to burn a carbon-fibre phenolic (CF-Ph) and a glass-fibre phenolic (GF-Ph) sample, both with integrated thermocouples, as well as a stainless-steel plate with welded thermocouples (Fig.~\ref{fig:samplecuts}) for \qty{5}{\minute}. After the test, square samples were cut from selected regions within the camera FOV to measure their averaged spectral emissivity on the back face with a reflectometer.The results for the three different materials and subregions (Fig.~\ref{fig:samplecuts}) are listed in Tab.~\ref{tab:emissivites} in the 'SOC100-HDR' column.

As a general trend for two composite materials, the emissivity is lower when CF is used as reinforcement compared to GF, and spatial variations are small (\textDelta\textepsilon{}~=~0.02). The average emissivity was 0.9 for GF-Ph samples, while it was 0.86 for CF-Ph samples. The effective emissivity of pristine CF samples has been measured as 0.88. Total emissivity of pristine GFRP is typically indicated between 0.9 and 0.95 in the literature \cite{Kim_Dembsey_2015, Berardi_Dembsey_2015}. As the resin burns the main contribution to the total sample emissivity shifts from the virgin matrix to the thermally degrading matrix and finally to the reinforcement fabric itself. For silica glass, emissivity values at room temperature are reported as 0.9-0.95 and for (graphitized) carbon 0.77-0.8\cite{Kreith2004}. Sample GF-Ph4 was essentially bare fabric following fire exposure and thus the value obtained of \textepsilon{}=0.91 is coherent with the literature data. 

For the large SAE 304 steel plate sample, the flame was impinging on the centre, approximately where SS-3 is located.  As Fig.~\ref{fig:samplecuts} shows, the thermal treatment of the steel resulted in an oxidation of the surface, which also increased its roughness. Away from direct flame impingement, the oxide layers create distinct “annealing colors” that give an estimate of the maximum temperature occurring in the zone. For instance, the golden yellow that transitions into purple as seen on sample SS-1 is representative for a temperature from \qtyrange{350}{450}{\degreeCelsius} for this alloy. The total emissivity measured for SS-3 (\textepsilon{} = 0.65) is in excellent agreement with values reported for comparable stainless-steel samples oxidized for \qty{5}{\minute} at \qty{800}{\degreeCelsius} \cite{Gordon2012}. The samples SS-1 and SS-3 are out of the FOV of the IR camera and cannot be compared against results from the TES algorithm. Based on literature data \cite{Evans1925, Higginson2015} the golden yellow oxide layer is estimated to be less than \qty{500}{\nm} and the blue oxide layer to be approximately \qty{200}{\nm} thick. Roebuck et al. \cite{Roebuck2013} have presented a study that shows that thin oxide layers like these can impose large temperature errors when using 1C pyrometry. The effect of the oxide layer thickness on the emissivity of steel, in particular for shorter wavelengths, has been previously reported to result in values coherent with our observations by Jo et al. \cite{Jo2017}. 

\begin{table}[hb]
\centering
\caption{Comparison of surface spectral emissivity measurements averaged over a square cutout of \qty{25.4}{\milli\metre} length after fire testing by hemispherical directional reflectometry (HDR) and 2C pyrometry. The sample numbering refers to Fig.~\ref{fig:samplecuts} and indicates the spatial location.}
\label{tab:emissivites}
\begin{tabular}{clccccc}
\toprule
Category & Sample & \multicolumn{5}{c}{Measured spectral emissivity, \textepsilon{}[-]}\\
\cmidrule{3-7}
    &   & \multicolumn{2}{c}{SOC100-HDR} & \multicolumn{2}{c}{IR camera} & Difference\\
    \cmidrule{3-4}
    \cmidrule{5-6}
    \cmidrule{7-7}
    &   &   Mean & SD & Mean & SD & \textDelta\textepsilon\\
    \midrule
    \multirow{6}{*}{Carbon-fibre phenolic} & CF-Ph1  & 0.88	& 0.02 &	0.65 & 0.04	&	0.23\\
    & CF-Ph2	& 0.86 &	0.02 &		0.70 &	0.04 &		0.15\\
    & CF-Ph3	& 0.86 &	0.02 &		0.65 &	0.04 &		0.21\\
    & CF-Ph4	& 0.86 &	0.02 &		0.70 &	0.04 &		0.16\\
    & CF-Ph5	& 0.86 &	0.02 &		0.76 &	0.04 &		0.11\\
    & mean	    & 0.86 &	  -   &     0.69 &	 -	 &      0.17 \\
    \midrule
\multirow{5}{*}{Glass-fibre phenolic} & GF-Ph1 &	0.90&	0.01	&	0.67 &	0.03	&	0.23\\
& GF-Ph2	& 0.88 & 	0.01	&	0.71 &	0.03 &		0.17\\
& GF-Ph3	& 0.89 &	0.01	&	0.63 &	0.03 &		0.26\\
& GF-Ph4	& 0.91 &	0.01	&	0.72 &	0.03 &		0.19\\
& mean	& 0.90 &	-		&   0.68 &	 -	 &  	0.22\\
\midrule
\multirow{3}{*}{Stainless steel} & SS-1 & 0.07 &	0.25	&	- &	-	&	- \\
 & SS-2	& 0.11 &	0.25	&	-    &	   -	&	-\\
 & SS-3	& 0.64 &	0.27    & 0.66	 &    -		& 0.02\\
\bottomrule
\end{tabular}
\end{table}

\subsection{Verification of gray body hypothesis}\label{sec:graybodyverif}

\begin{figure}[ht]
    \centering
    \begin{subfigure}[t]{0.45\textwidth}
    \centering
         \includegraphics[width=0.9\textwidth]{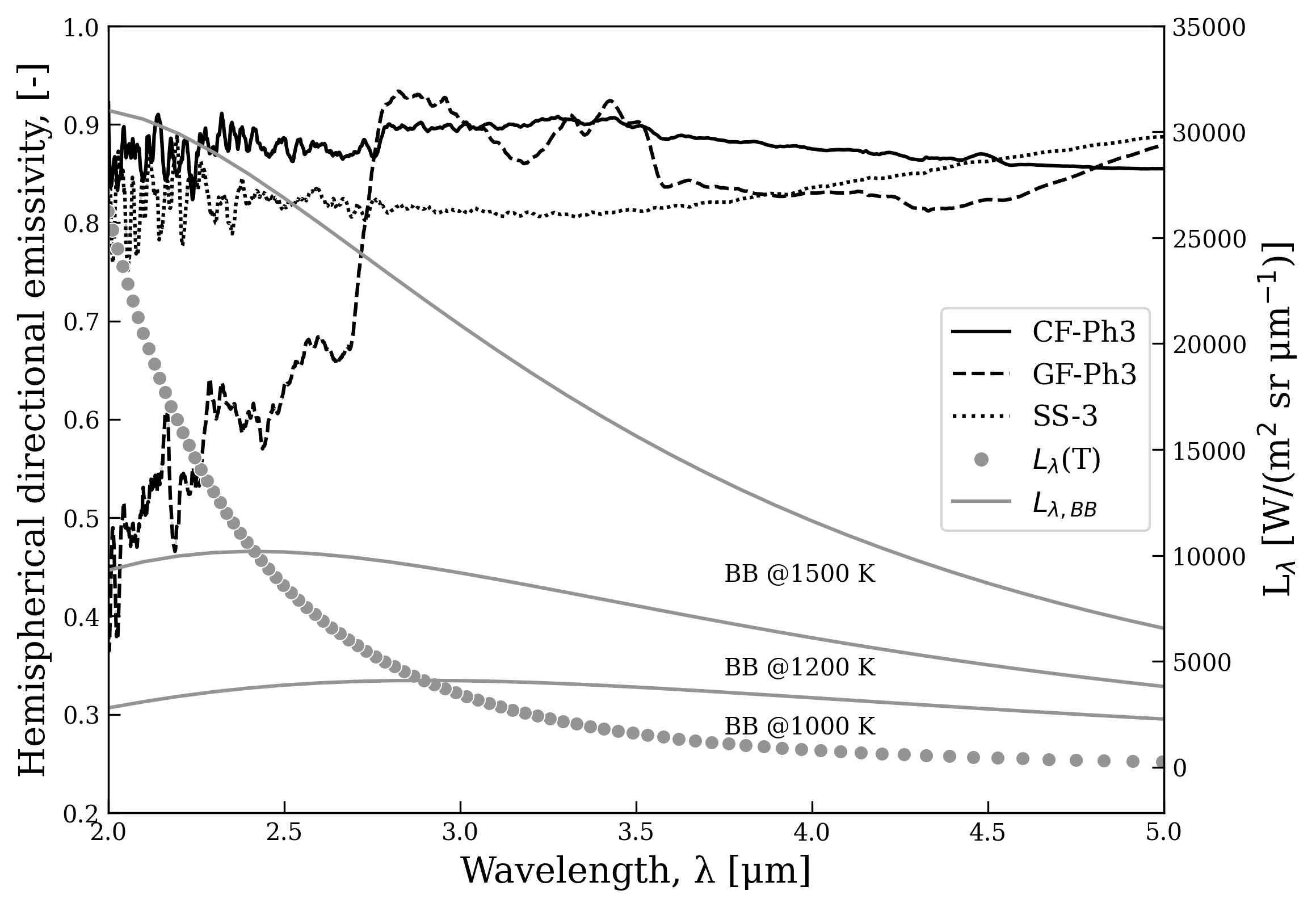}
    \caption{}
    \label{fig:hdr1}
    \end{subfigure}
    \begin{subfigure}[t]{0.45\textwidth}
    \centering
         \includegraphics[width=0.85\textwidth]{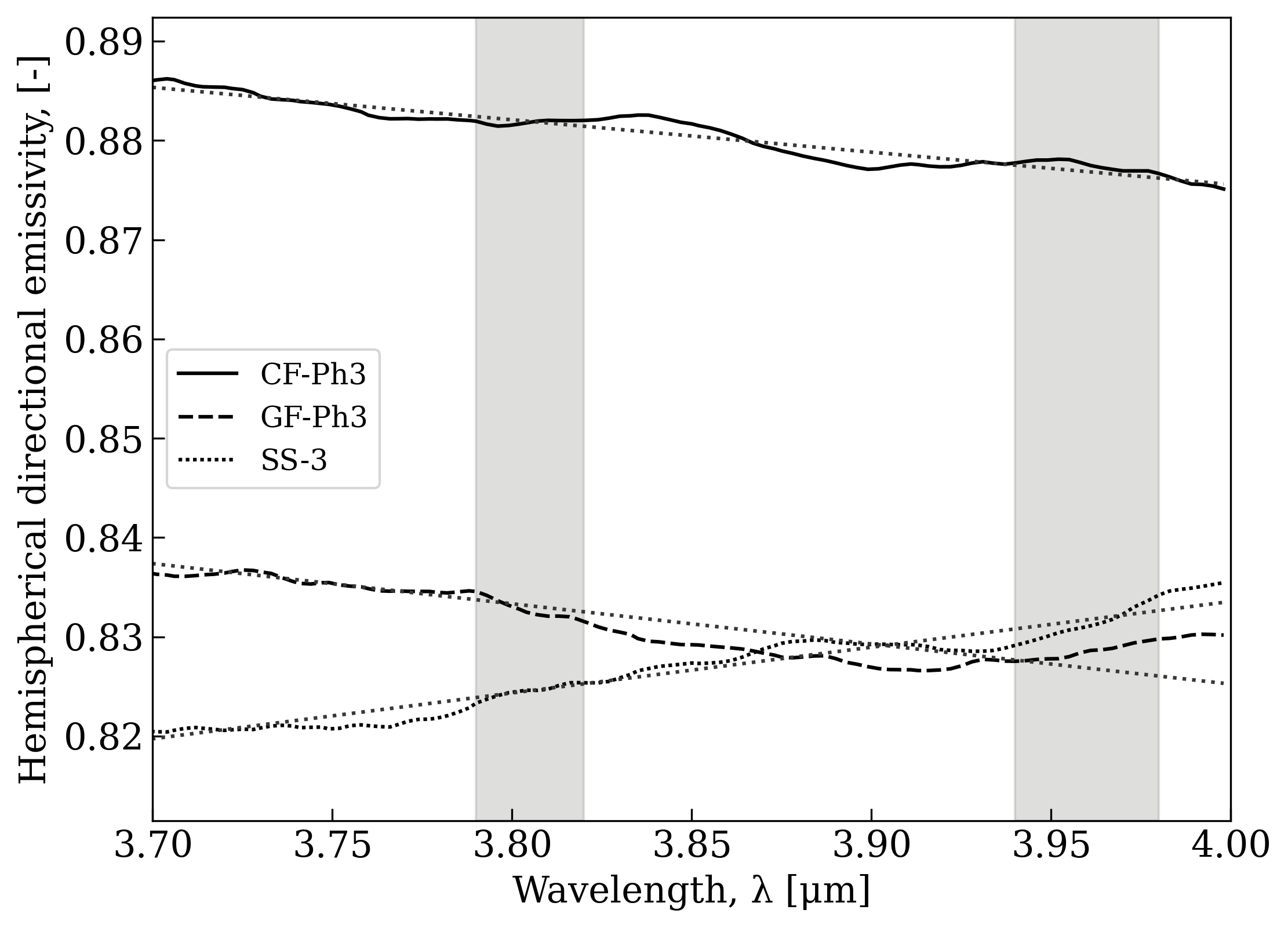}
    \caption{}
    \label{fig:hdr2}
    \end{subfigure}
        \caption{(a) Hemispherical directional emissivity of a carbon-fiber phenolic (CF-Ph), glass-fiber phenolic (GF-Ph) and stainless-steel (SS) sample degraded under a kerosene flame of 116 kW/m2. The radiance of a black-body at different temperatures is overlaid. (b) Emissivity in the spectral region of the filters used for 2C pyrometry, shown as gray bars.}
    \label{fig:hdr}
\end{figure}

The HDR results presented in Sec.~\ref{sec:emHDR} and in Tab.~\ref{tab:emissivites} give the effective emissivites of the samples, that is the integrated value over the entire spectral range. However, emissivities often show a spectral dependency, as shown in  Fig.~\ref{fig:hdr1} for three different samples. The thermally degraded GF-Ph3 sample showed a pronounced decrease of emissivity at wavelengths shorter than \qty{2.5}{\micro\meter}. The emissivity of the CF-Ph3 and steel SS-3 samples remained relatively constant over the spectral range used in the HDR measurements. Two important assumptions of our radiation model are that Kirchhoff’s law can be applied (cf. to Tab.~\ref{tab:rmodel}) and that the sample emissivity is constant within the spectral band of the filters. The HDR measurements were used to test this last hypothesis and the hemispherical emissivity data shown in Fig.~\ref{fig:hdr2} is linearly fitted. Although the slope is different from unity for all materials, the emissivity varies not more than \qty{0.14}{\percent} for all samples over the spectral bands of filter FW\#7 and FW\#8. It is thus reasonable to assume constant values within each filter band. Nevertheless, the emissivity may vary from one filter to another by a conservative estimate of \qty{1}{\percent} at 0.8 nominal emissivity. The resulting error for a given filter contributes with \qty{8.1}{\degreeCelsius} to the overall error budget. 

\subsection{Insitu emissivity measurements}\label{sec:insitu-em}

\begin{figure}[tb]
    \centering
    \begin{subfigure}[t]{0.45\textwidth}
    \centering
         \includegraphics[width=\textwidth]{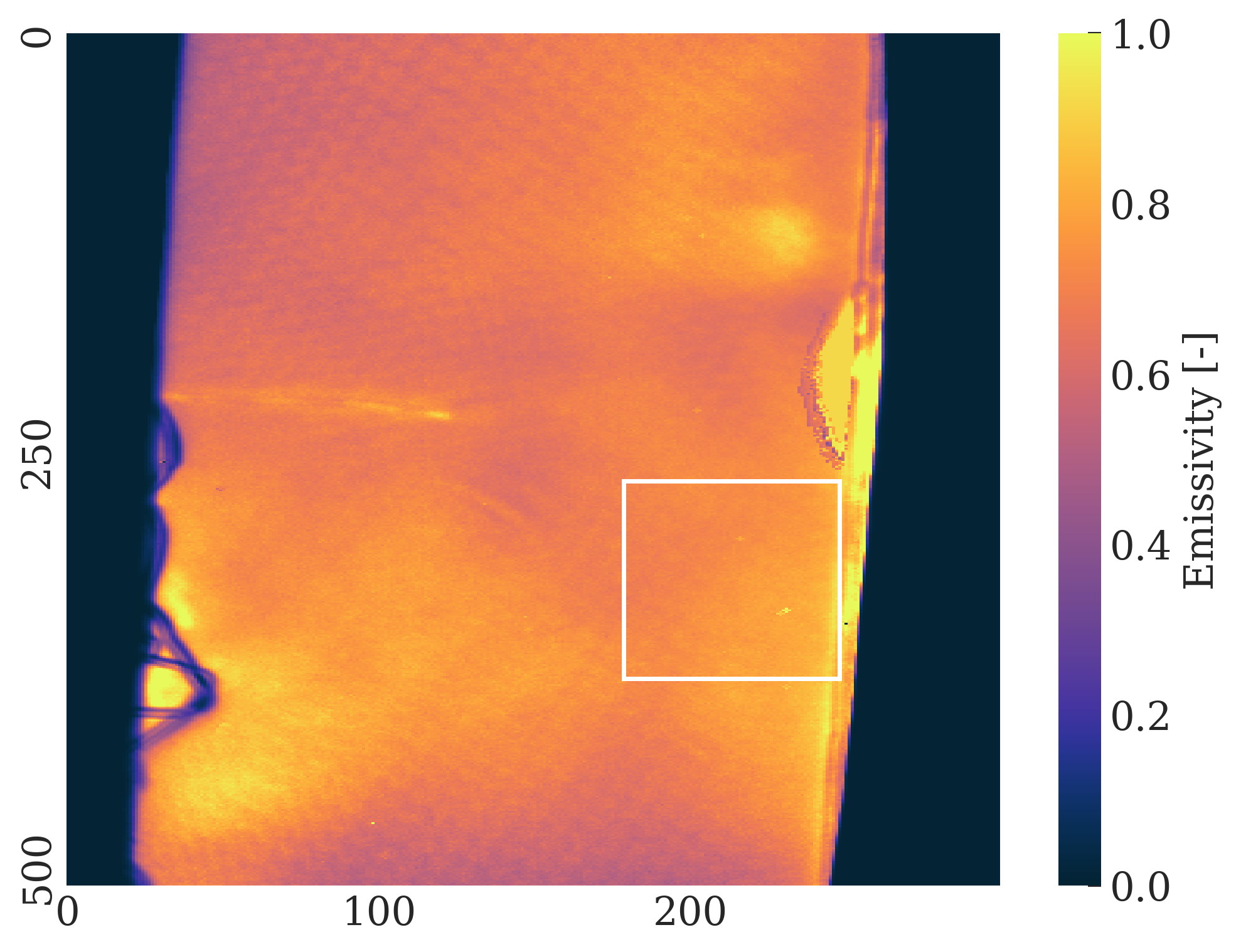}
    \caption{}
    \label{fig:CFeall}
    \end{subfigure}
    \begin{subfigure}[t]{0.45\textwidth}
    \centering
         \includegraphics[width=\textwidth]{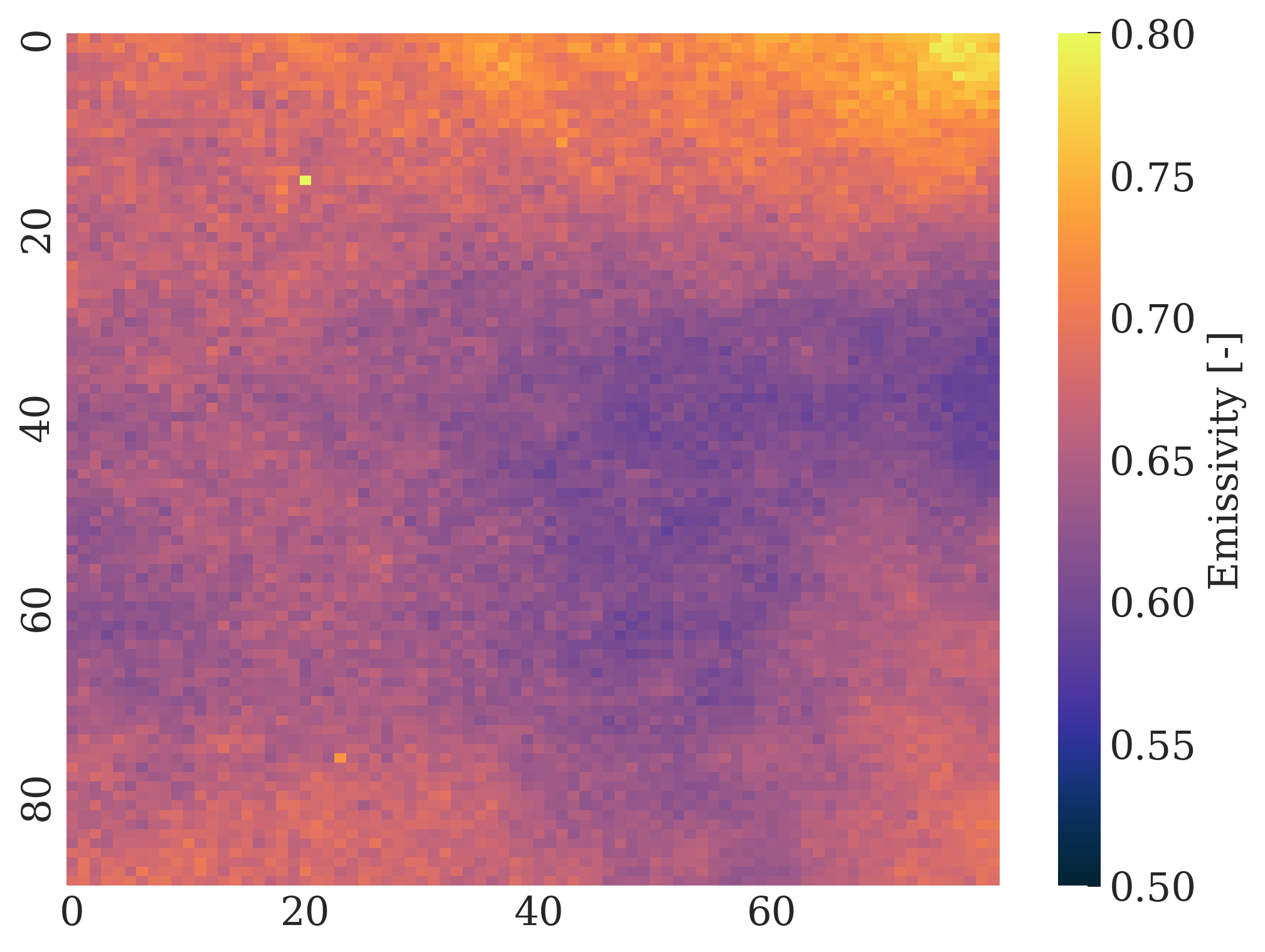}
    \caption{}
    \label{fig:CFesub}
    \end{subfigure}
    
        \begin{subfigure}[t]{0.45\textwidth}
    \centering
         \includegraphics[width=\textwidth]{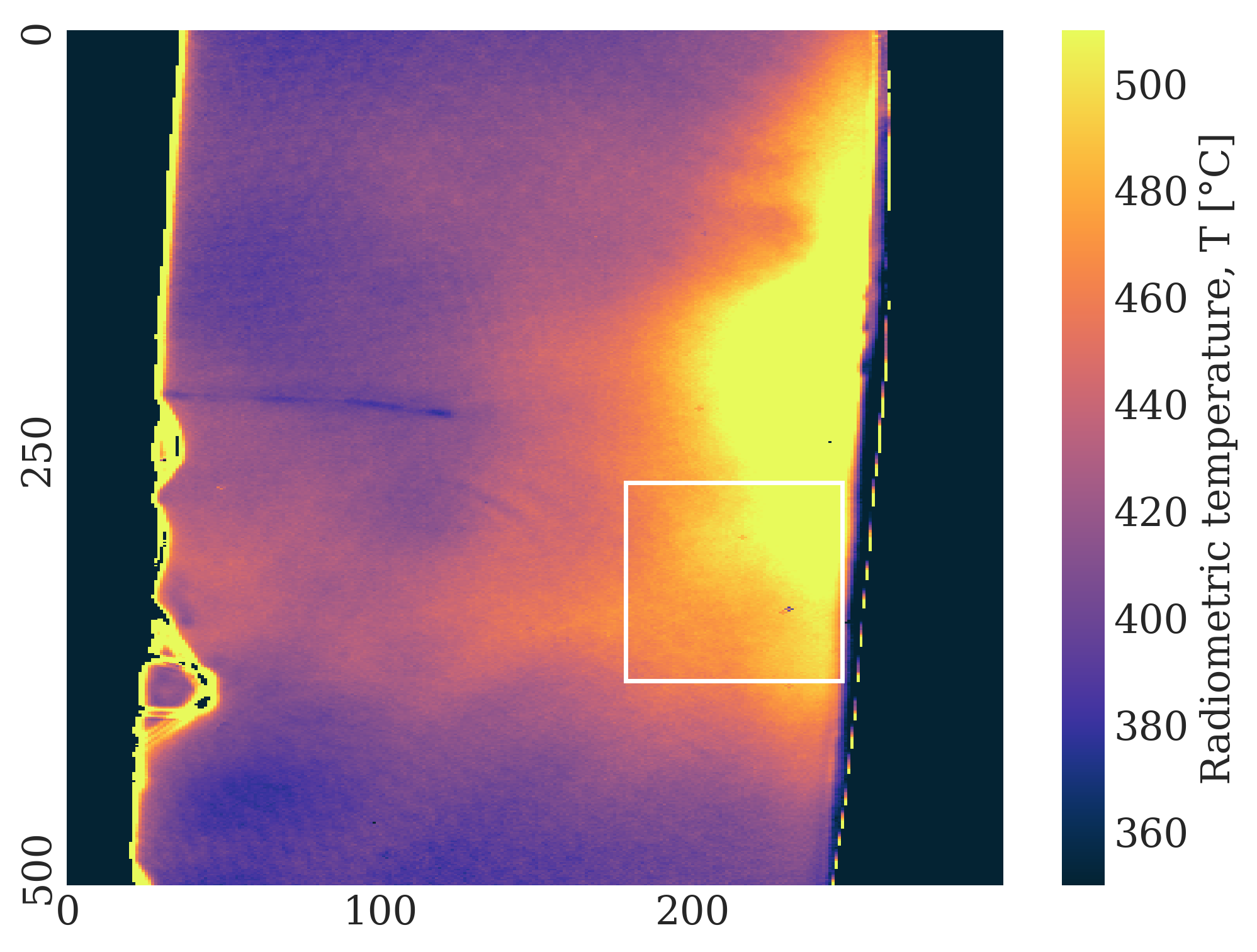}
    \caption{}
    \label{fig:CFRTall}
    \end{subfigure}
    \begin{subfigure}[t]{0.45\textwidth}
    \centering
         \includegraphics[width=\textwidth]{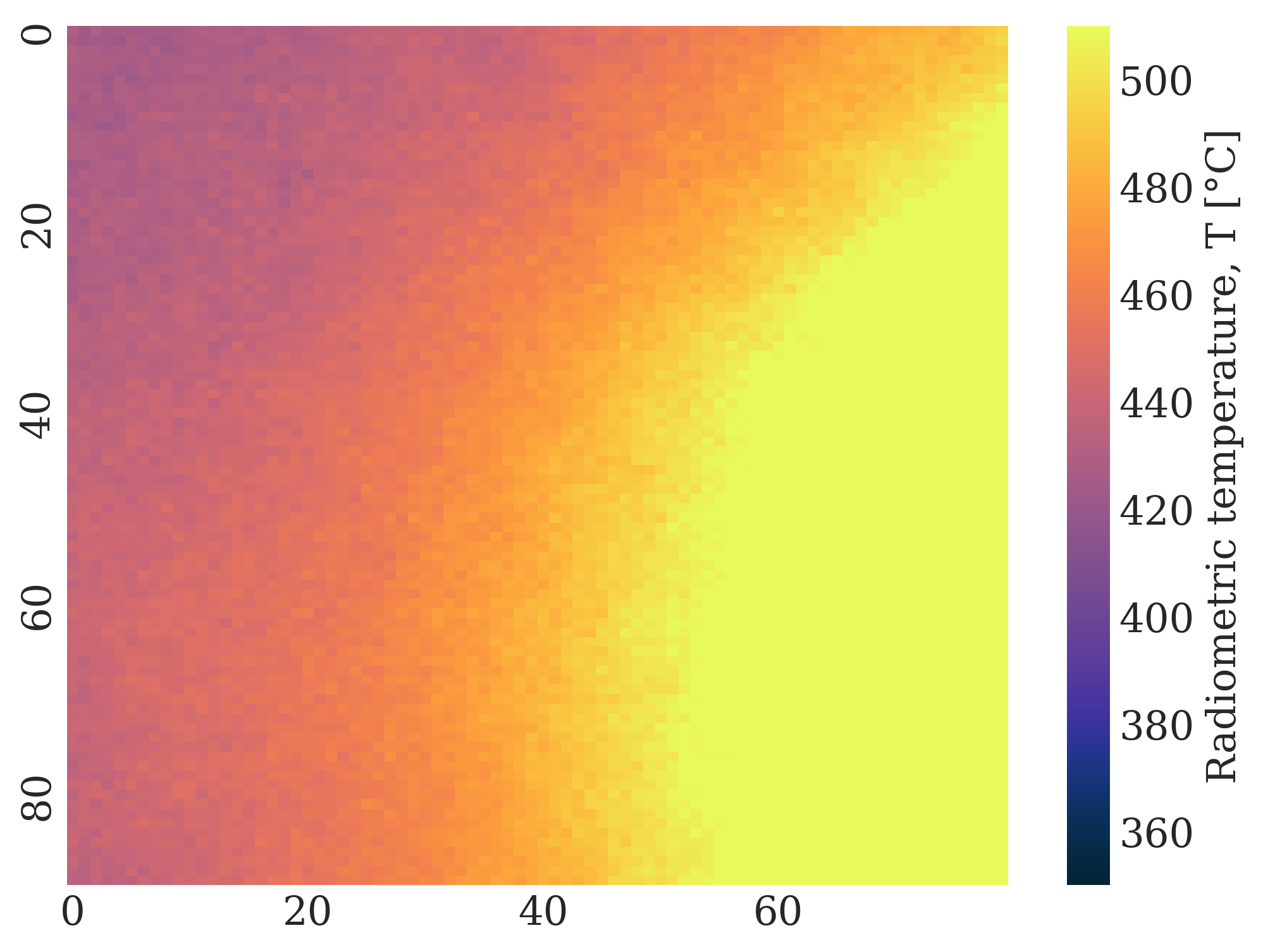}
    \caption{}
    \label{fig:CFRTsub}
    \end{subfigure}
        \caption{Emissivity and temperature mapping of a CF-epoxy sample exposed to a flame. The images on the right are the subregions indicated by the white rectangle. The subregions are used to calculate average value. The colorbar range for the emissivity subrange has been modified to better illustrate spatial variations.}
    \label{fig:emissTemp-CF}
\end{figure}

Comparison between the emissivity mapping obtained from the TES algorithm and the HDR measurements was carried out by computing the average over the pixel-area within the FOV that corresponds to the sample cutouts in Fig.~\ref{fig:samplecuts}. Two important points must be highlighted for the subsequent discussion. First, the HDR measurements were performed at room temperature following fire exposure. Second, the minimum radiance level necessary to overcome the noise level for the filters used require minimum sample temperature of approximately \qty{300}{\degreeCelsius}. In addition to this temperature difference, the sample also continues to degrade during cool down and thus the sample state as seen by the HDR can differ from the end of the sequence acquired by MS-IR camera. To minimize this discrepancy, only the last 50 frames of the IR acquisition were used to calculate emissivity and temperature maps of the sample subregions. The results from this analysis are summarized in Tab.~\ref{tab:emissivites}. Representative emissivity and temperature maps for the CF-Ph sample at approximately \qty{380}{\degreeCelsius} are shown in Fig.~\ref{fig:emissTemp-CF}, together with the subregions that correspond to sample CF-Ph3. 

It is apparent from Tab.~\ref{tab:emissivites} that the insitu emissivities obtained with the TES algorithm are significantly lower for the composite samples than the ones analysed post-fire. This is a caused by the complex temperature dependency of the emissivity. The values we report are consistent with a similar CFRP sample subjected to the same burner and conditions, where TES analysis yielded comparable values in the range of 0.5-0.7 \cite{Gnessougou2019}. Althoug Fig.~\ref{fig:emissTemp-CF} illustrates that the thermal degradation of composite materials can display large spatial variations, an averaged emissivity value is still meaningful if the ROI is chosen accordingly. 

Jones et al. \cite{Jones_Mason_Williams_2019} proposed a model for the temperature dependence of glass-fibre composites emissivity. Using their model and accounting for the temperature difference (\textDelta T $\approx$ \qty{420}{\degreeCelsius}) between HDR and TES measurements, the coefficients identified from GF-Ph data corresponds well to the coefficients typical for glasses. It is to be expected as resin depletion at the end of the test results in the exposed surface being mostly reinforcement fibres. Furthermore, the decrease in emissivity from approximately 0.9 at room temperature (HDR at \qty{20}{\degreeCelsius}) to 0.65 at \qty{440}{\degreeCelsius} (TES with \textDelta T = \qty{420}{\degreeCelsius}) is also consistent with the high temperature emissivity measurements of quartz glass at similar temperatures reported in the literature \cite{Petrov_Reznik_1972, Barnes_Forsythe_Adams_1947}. 

For the stainless-steel sample, the emissivity obtained for SS-3 through TES calculations from IR data is in excellent agreement with the HDR measurements. Once the oxide layer has reached substantial thickness (\textgreater \qty{10}{\micro\meter}), temperature and emissivity variations over successive 50 frames interval can be neglected.  Moreover, the oxide layer reduces reflective contributions which contributes to reduce the overall error. 

The results from this analysis suggest that total emissivity can indeed be regarded as independent from wavelength within the spectral range of the filters. Moreover, it can be justified to use averaged emissivity values for subregions that are homogeneously burnt. A global constant emissivity value measured at room temperature is not representative for high-temperature tests for composite samples not even for ones that are significantly degraded and whose emissivity is mainly determined by their fibre reinforcement. Although the emissivity from steel samples has a less pronounced temperature dependence, it is important to notice that the surface oxidation progresses rapidly under flame exposure resulting in drastically changed emissivity and making it extremely challenging to measure surface temperatures during heat-up with conventional IR thermography. 

\subsection{Insitu temperature measurements}

\begin{figure}[tb]
     \centering
    \begin{subfigure}[t]{0.45\textwidth}
    \centering
         \includegraphics[width=.9\textwidth]{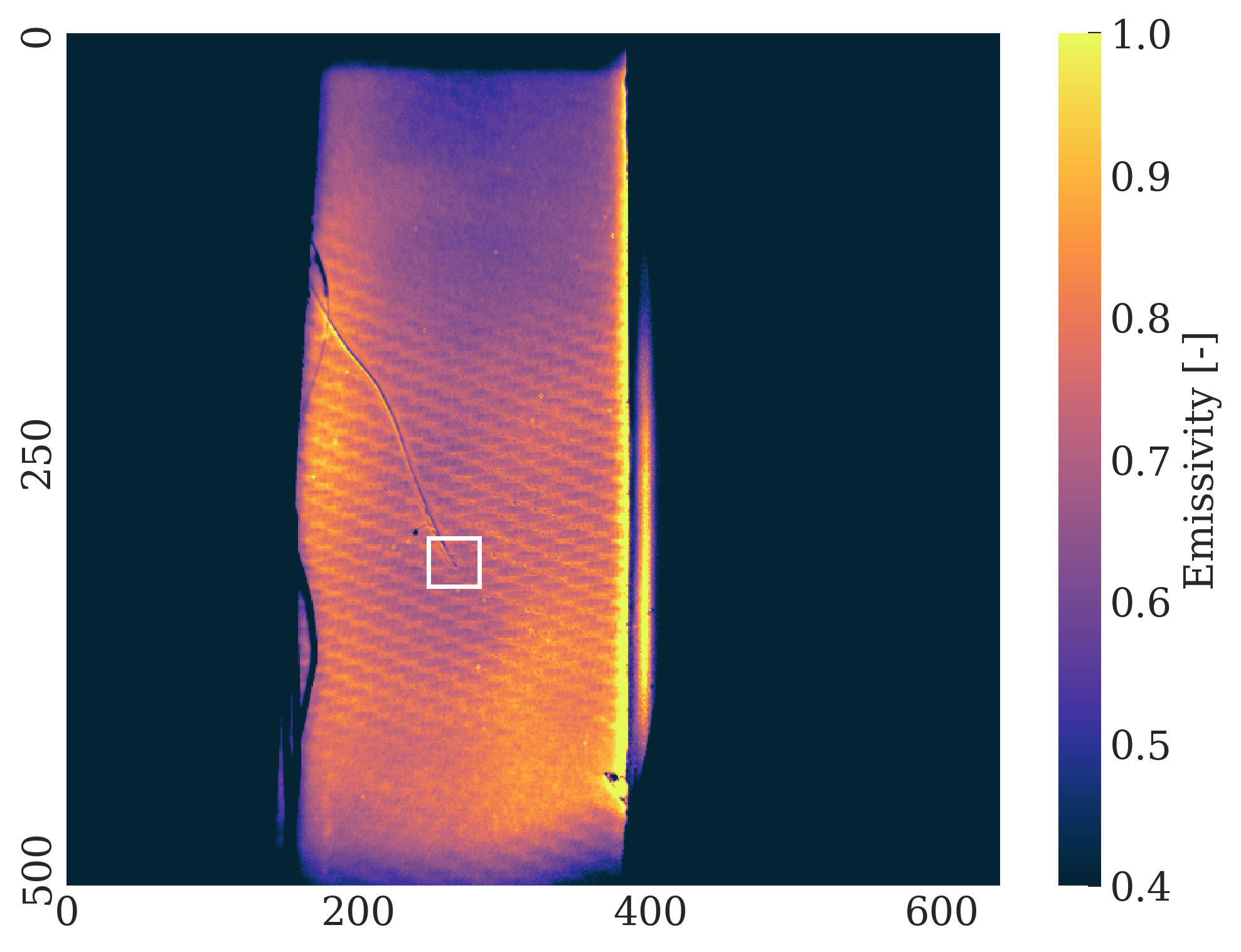}
    \caption{}
    \label{fig:CFepall}
    \end{subfigure}
    \begin{subfigure}[t]{0.45\textwidth}
    \centering
         \includegraphics[width=\textwidth]{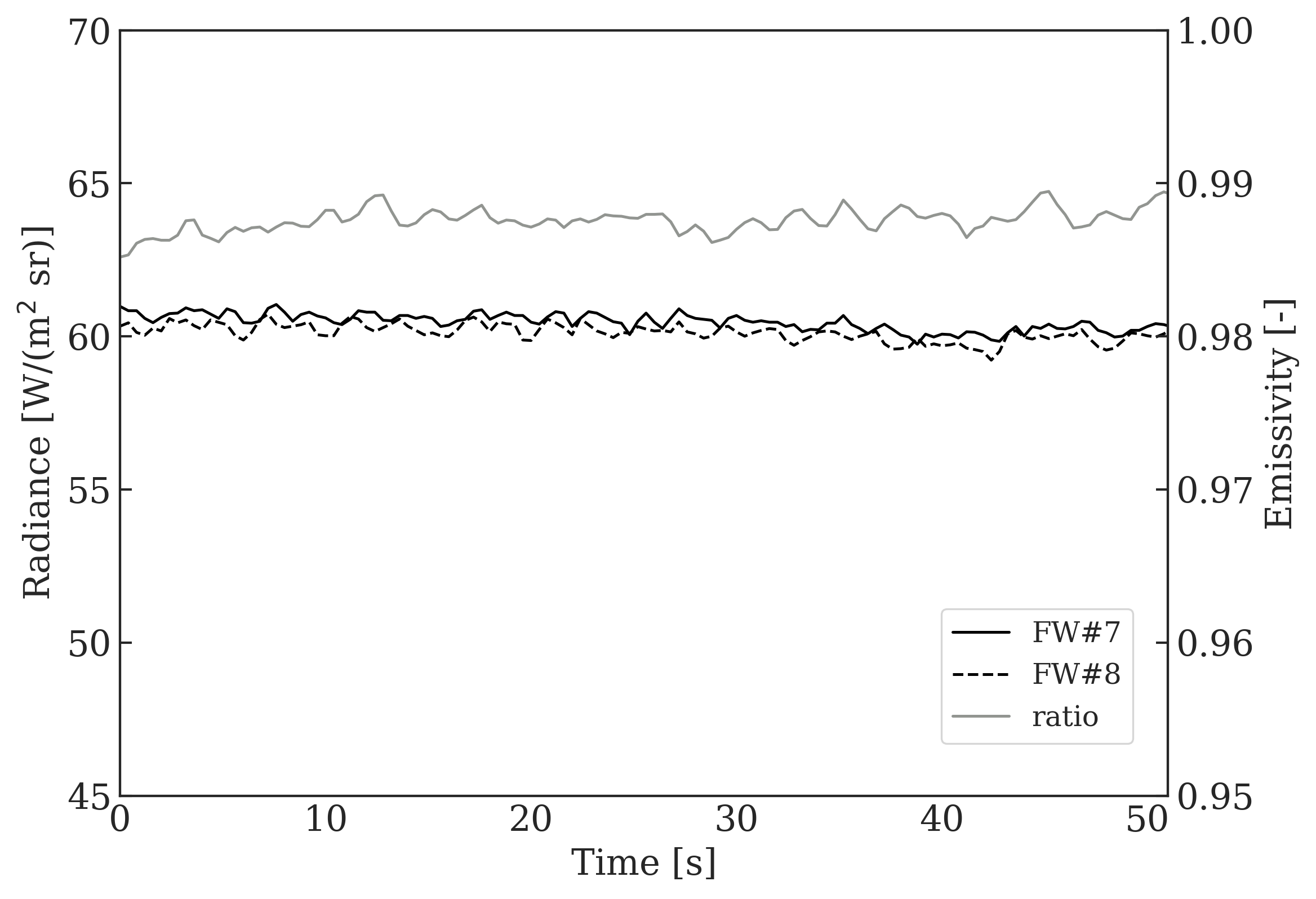}
    \caption{}
    \label{fig:CFesignal}
    \end{subfigure}
    \caption{(a) IR image of the sample backside, with the attached thermocouple wire clearly visible. A measurement area of \qtyproduct{20 x 20}{\pixel} has been placed around the thermocouple junction (white rectangle). (b) Raw radiance signal from the two filter used to calculate the IBR-ratio.}
    \label{fig:proofofconcept}
\end{figure}

To test the radiometric accuracy of the TES algorithm, the temperature was simultaneously acquired at the back of a CF epoxy sample using a thermocouple probe and the IR camera. The acquisition frame rate was set to \qty{20}{\Hz} as temperature changes were not expected to be fast, based on previous TC measurements on similar samples. Fig.~\ref{fig:proofofconcept} shows an example of an infrared image of the sample back face during cool-down, with the integrated thermocouple clearly visible. During post-processing, an area of \qtyproduct{20 x 20}{\pixel} at the location of the thermocouple probe is used to obtain an average signals for the filters ( Fig.~\ref{fig:proofofconcept}).  The raw signals (Fig.~\ref{fig:CFesignal}) from FW\#7 and FW\#8 are smoothed using a moving average and then used to form the signal ratio. In a final step the temperature and emissivity of the sample in the vicinity of the thermocouple is calculated from the IBR ratio. The final IR temperature signal was further smoothed with a Savitzky-Golay filter with a window size of 69 successive frames.

\begin{figure}[ht]
    \centering
    \begin{subfigure}[t]{0.45\textwidth}
    \centering
         \includegraphics[width=\textwidth]{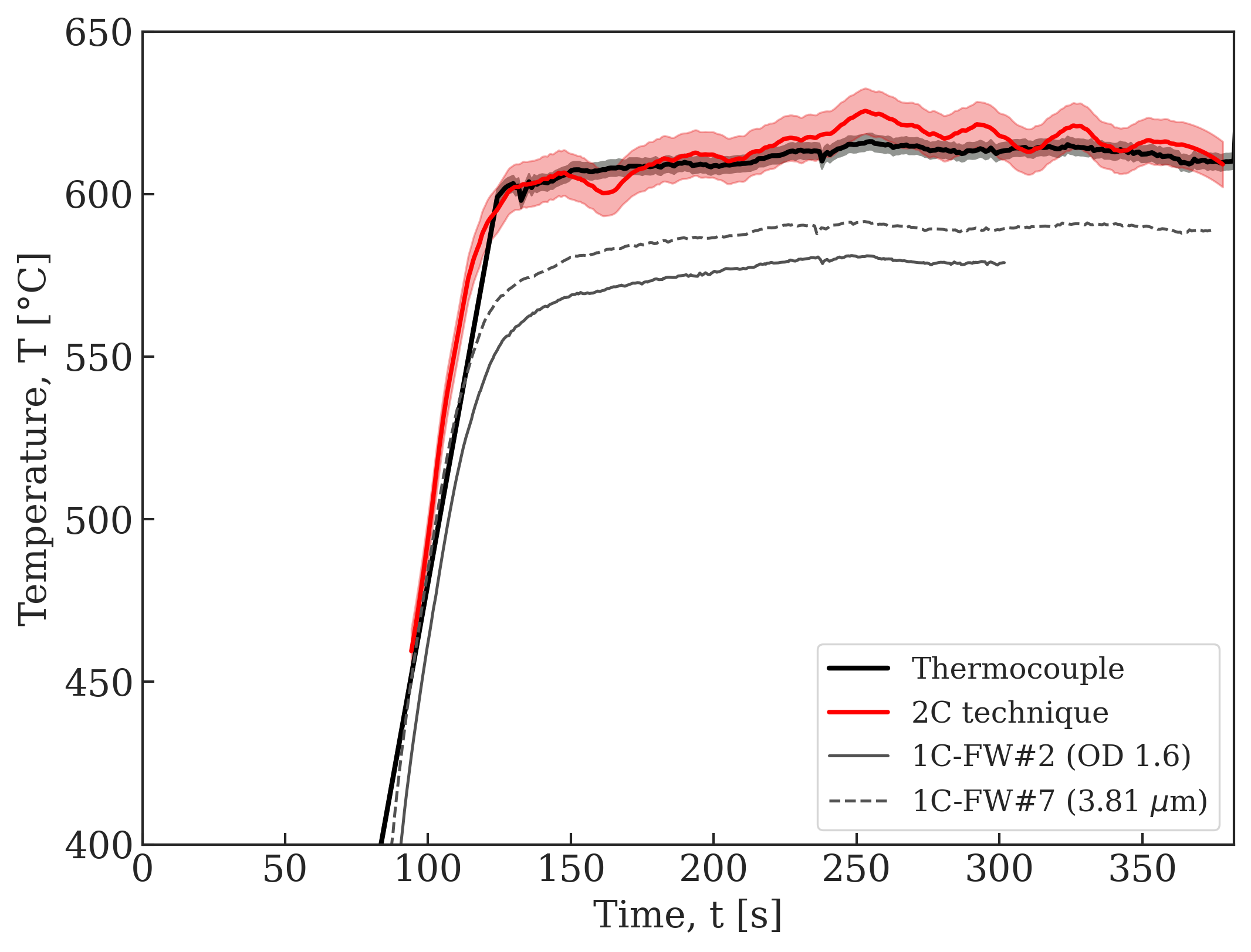}
    \caption{}
    \label{fig:final-T}
    \end{subfigure}
    \begin{subfigure}[t]{0.45\textwidth}
    \centering
         \includegraphics[width=\textwidth]{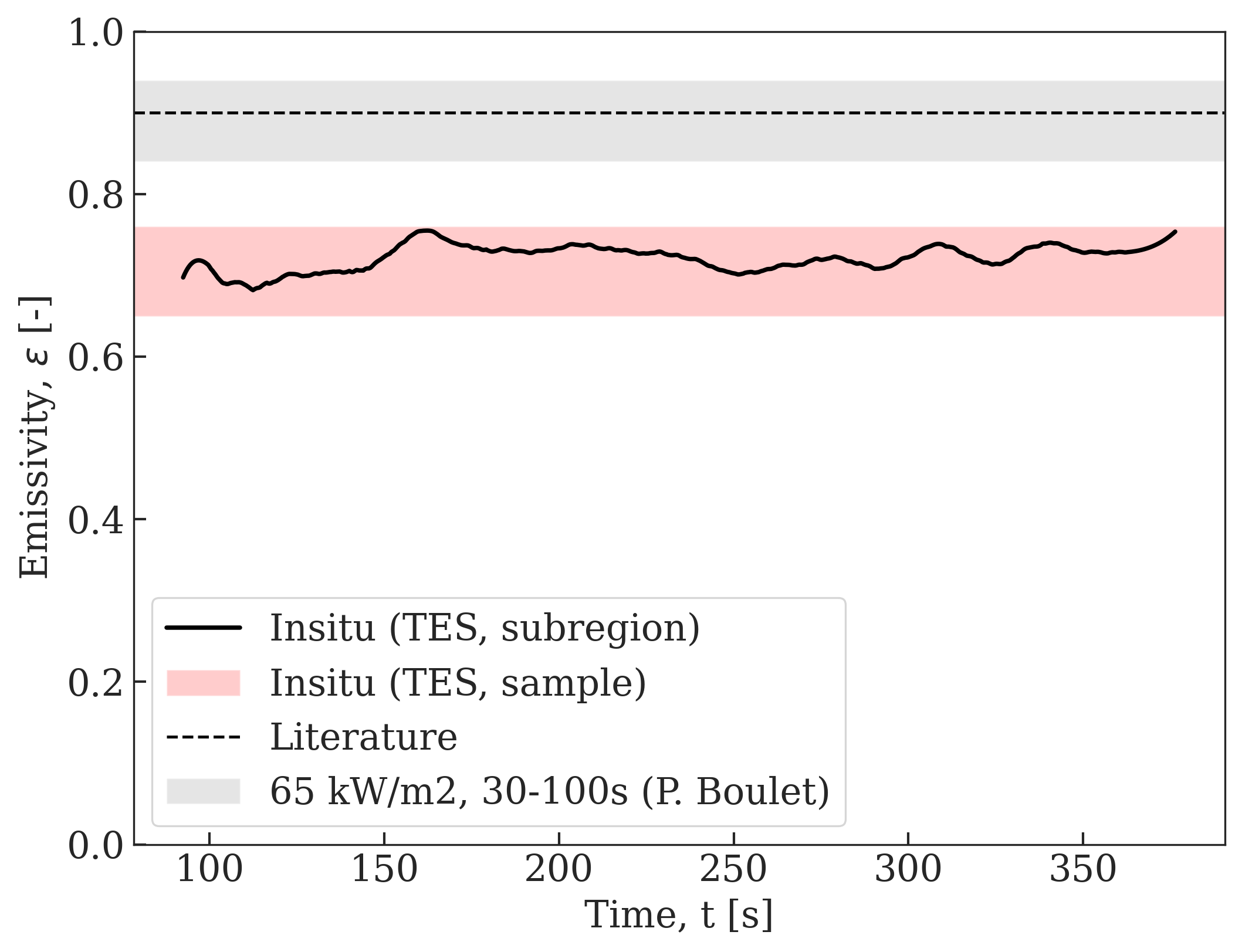}
    \caption{}
    \label{fig:final-e}
    \end{subfigure}
        \caption{(a) Thermocouple signal compared against the temperature signal retrieved using the 2C technique from a \qty{20}{\Hz} frame rate acquisition as well as 1C thermography calculations. (b) Emissivity evolution on the backface of CF composite sample exposed to a flame measured through 2C pyrometry. The gray area corresponds to the emissivity measured for different zones on a similar sample subjected to comparable test conditions from literature \cite{Boulet_Brissinger_Collin_Acem_Parent_2015}.}
    \label{fig:finalRT-e}
\end{figure}

Fig.~\ref{fig:final-T} shows the temperature curves obtained through the TES alogorithm compared against the TC measurements. The acquisition with the IR camera was triggered slightly delayed so that the first data points start when the back face temperature has already risen above \qty{400}{\degreeCelsius}. The curve obtained with the 2C technique implementing the necessary corrections is in excellent agreement with the thermocouple data. 
The results are compared against an implementation of conventional 1C pyrometry data, using only a neutral density filter (FW\#2) or a single channel (FW\#7), relying on the emissivity of pristine CF epoxy samples (\textepsilon{}=0.88).  This data illustrates that assuming a constant emissivity value for composite samples, often taken from post-fire measurements, might result in high errors in temperature readings, as variation exceeding \qty{10}{\percent} are easily observed during the test, especially when doing single-color pyrometry. 

With the temperature available, the emissivity evolution was calculated and is shown in Fig.~\ref{fig:final-e} for approximately the last five minutes of a fire exposure test. The emissivity range obtained for the CF-epoxy samples averaged over different subregions through the TES algorithm is red shaded. The gray shaded area represents literature data for the emissivity of CF-epoxy samples exposed from \qtyrange{30}{100}{\second} to a radiant heat flux of \qty[per-mode=symbol]{65}{\kilo\watt\per\square\metre}, with the dashed black line indicating an emissivity value often used for 1C pyrometry measurements in similar fire exposure tests \cite{Schuhler_Chaudhary_Vieille_Coppalle_2021, Tranchard_2015}. 

\subsection{Error budget}

Tab.~\ref{tab:budget} lists the individual contributions to the overall error budget, along with typical values. 
The temporal noise is the uncertainty of measurement associated to a single temperature reading and thus a single pixel. The radiance ratio suffers from the temporal noise in each of the two filter channels. The typical FW\#7 camera noise for an apparent temperature of \qty{500}{\degreeCelsius} is \qty[per-mode=symbol]{30}{\milli\watt\per\square\metre\per\steradian} for a radiance of approximately \qty[per-mode=symbol]{43.5}{\watt\per\square\metre\per\steradian}. This figure of performance is similar with the second filter. The resulting standard deviation on the radiance ratio is then 0.00094, translating to an error of \qty{3.8}{\degreeCelsius} for a scene of approximately \qty{500}{\degreeCelsius}. This pixel-wise performance figure can be improved with either temporal or spatial averaging. The temporal noise measurement for high temperature targets can be for instance be performed with a high temperature black-body in front of the camera, but is difficult due to thermal turbulence.  Therefore, the temporal noise is simulated and scaled to the filter response to provide the noise equivalent differential IBR. 

If the image superposition is not perfect between the two frames used for the 2C pyrometry method, a parallax error is associated with a band ratio difference, leading to inaccurate temperature-emissivity retrievals. The camera enables a position correction for each spectral band in both axes, but results on the sample backside did not require its implementation for our tests. Since the spatial resolution is about \qty{0.6}{\milli\metre} per pixel, the region of interest is chosen wide enough to spatially average the signal. This allows for an improved signal to noise ratio for the macro-pixel. The error caused by the difference in spatial features from one filter to the other is evaluated to be lower than \qty[per-mode=symbol]{0.01}{\watt\per\square\metre\per\steradian}.

Radiometric accuracy is verified against a calibrated black-body for each spectral band. The camera absolute accuracy is found to be in the order of \qty{\pm 0.2}{\percent} for a black-body set at \qty{500}{\degreeCelsius}, that translates into an error contribution of \qty{5.5}{\degreeCelsius} for the retrieved temperature. 

As outlined in Sec.~\ref{sec:rc}, the filter-wise radiometric correction requires an estimation of the number of pixels in the illumination area. The uncertainty of providing the accurate illumination area leads to a maximum error of \qty{0.7}{\percent} in the illumination ratio for each spectral band. The resulting error is \qty{6.4}{\degreeCelsius} for a scene at \qty{500}{\degreeCelsius}.

The assumption that the emissivity remains constant within the filter spectral range for both filters also contributes to the error budget. For a maximum error of \qty{1}{\percent} in the emissivity difference between the two filter bands, the resulting error is \qty{8.1}{\degreeCelsius}. 

For the measurements considered here where both the sample temperature and emissivity change with time, the choice of the acquisition frequency can be an important source of error.  Ideally, the frame rate must ideally be selected such that the overall uncertainty of the measurement is on the same order of magnitude as the temporal noise. The temperature increase  remained below \qty[per-mode=symbol]{0.224}{\degreeCelsius\per\second} at a sample temperature of approximately \qty{500}{\degreeCelsius}. Acquiring the signal at a frequency of \qty{0.8}{\Hz} allowed to keep the maximum error below 
\qty{0.28}{\degreeCelsius}. The corresponding error contribution is \qty{7}{\degreeCelsius}. However, if the temperature increase is instead \qty[per-mode=symbol]{2.24}{\degreeCelsius\per\second}, the associated error is ten times higher. This specific behaviour underlines why great care must be applied to select a frame rate sufficiently high to limit frame rate related errors as they can rapidly overwhelm other sources of uncertainty.

\begin{table}[hb]
\centering
\caption{Example of a typical error budget for a measured sample temperature of \qty[separate-uncertainty=true]{500 \pm 28}{\degreeCelsius}. The reported expanded uncertainty U of measurement is stated as the standard uncertainty of measurement (u) multiplied by the coverage factor k = 2, which for a normal distribution corresponds to a coverage probability of approximately \qty{95}{\percent}. }\label{tab:budget}
\begin{tabular}{llllc}
\toprule
\multicolumn{2}{c}{Quantity}  &	\multicolumn{2}{c}{Maximum error}   &	 Uncertainty\\
    \cmidrule{1-2}
    \cmidrule{3-4}
    \cmidrule{5-5}
 Source& Description &  Estimate & u & [\unit{\degreeCelsius}] \\ 
\midrule
\multirow{2}{*}{Pixel-wise} & Temporal noise & \qty{0}{\mW/(\m^2 \steradian)} & \qty{30}{\mW/(\m^2 \steradian)} &  3.8 \\
& Parallax error & \qty{<60}{\mW/(\m^2 \steradian)}  & \qty{10}{\mW /(\m^2 \steradian)} &  0.9\\
\midrule
\multirow{4}{*}{Filter-wise} & 
Camera accuracy  & \qty{<0.20}{\percent} & \qty{10.00}{\percent} & 5.5 \\
& Geometric correction &	\qty{<0.70}{\percent} 	&  \qty{0.12}{\percent}&  6.4 \\
&  Emissivity assumption &	\qty{<1.25}{\percent}  &	\qty{0.21}{\percent} &  8.1 \\
& Frame rate selection  &	\qty{<0.224}{\degreeCelsius/\s} &	\qty{0.224}{\degreeCelsius/\s} & 7.0 \\
\midrule
& & &  total & 14.2 \\
& & &  expanded uncertainty U& 28.4 \\
\bottomrule
\end{tabular}
\end{table}

\section{Conclusion}
Infrared thermography is a powerful tool for in-situ observation of the temperature evolution of fast degrading materials, such as polymer-matrix composite samples exposed to a flame. However, conventional IR thermography requires the prior knowledge of the sample emissivity. As burning composites degrade rapidly and the metallic samples carry over large contributions from the surroundings, the proper characterization of the source emissivity becomes virtual impossible. Two-color pyrometry is a technique that can overcome this problem, but although it becomes popular in the fire testing community its common error sources are often unappreciated. In this work, the temperature and emissivity evolution on the back face of samples subjected to a flame were studied using a multispectral IR camera equipped with a fast-rotating filter wheel. 

We present a detailed radiometric model to account for the contributing factors of the overall radiance signal originating from the sample itself, the background and the atmosphere. The background contribution is particularly significant below \qty{300}{\degreeCelsius} but can be neglected in our case as the optical band-pass filter selected does not detect below this temperature. A thorough sensitivity analysis demonstrated that two basic assumptions must hold true between two subsequent acquisition frames: first, the material  under thermal degradation must behave as a gray-body and second, the transmission of the selected optical filters must not depend on the emissivity of the object within their respective spectral band. 

We implement a multi-step image analysis process, associated experiments to obtain necessary material properties. In a first step, a geometric correction helps to adapt the camera calibration to the actual illuminated area and thus to take into account only the pixels that contribute to the overall radiance signal. Without such a correction, the target temperature was underestimated by \qty{10}{\degreeCelsius} and yielded unphysical emissivity values above one. In a second step, we subjected carbon-fiber and glass-fiber composites as well as stainless steel plates to a calibrated kerosene flame. The emissivity of the burnt samples at different position was also measured using hemispherical directional reflectometer. It could be shown that these composite materials exhibit the same selective transmitters tendency as other polymer and that their emissivity decreases as a function of temperature, relative to measurements conducted at room temperature. For the steel sample tested, the thickness of the oxidation layer is important and the spatial variation of emissivity ranged from \qtyrange{0.07}{0.64}{} for heavily oxidized steel plates  . 

The emissivity results obtained using the 2C technique and the proposed corrections are in excellent agreement with literature data measurements obtained from embedded thermocouples. The experimental methods, hardware selection and image post-processing including necessary corrections presented here allows the measurement of the back face temperature of composite materials exposed to a flame within an uncertainty below \qty{30}{\degreeCelsius} at a temperature of \qty{500}{\degreeCelsius}. 

\section*{Acknowledgements}
The authors wish to thank Jean-Philippe Gagnon for his expertise and valuable support during the initial experimental campaign.

\bibliographystyle{unsrt}  
\bibliography{references} 

\begin{thebibliography}{10}

\bibitem{Lautenberger_Fernandez-Pello_2009}
Chris Lautenberger and Carlos Fernandez-Pello.
\newblock Generalized pyrolysis model for combustible solids.
\newblock {\em Fire Safety Journal}, 44(6):819–839, Aug 2009.

\bibitem{Baukal_Gebhart_1996}
C.~E. Baukal and B.~Gebhart.
\newblock A review of empirical flame impingement heat transfer correlations.
\newblock {\em International Journal of Heat and Fluid Flow}, 17(4):386–396,
  1996.

\bibitem{Baukal_Gebhart_1995}
C.~E. Baukal and B.~Gebhart.
\newblock A review of flame impingement heat transfer studies part 1:
  Experimental conditions.
\newblock {\em Combustion Science and Technology}, 104(4–6):339–357, 1995.

\bibitem{Walter2006}
Géza Walter, László~I. Kiss, André Charette, and Vincent Goutière.
\newblock Flame-object heat transfer using different burner types and
  orientations.
\newblock {\em TMS Light Metals}, 2006(December 2014):741–746, 2006.

\bibitem{Bal_Rein_2013}
Nicolas Bal and Guillermo Rein.
\newblock Relevant model complexity for non-charring polymer pyrolysis.
\newblock {\em Fire Safety Journal}, 61:36–44, 2013.

\bibitem{Stoliarov_Crowley_Lyon_Linteris_2009}
Stanislav~I. Stoliarov, Sean Crowley, Richard~E. Lyon, and Gregory~T. Linteris.
\newblock Prediction of the burning rates of non-charring polymers.
\newblock {\em Combustion and Flame}, 156(5):1068–1083, May 2009.

\bibitem{Forsth_Roos_2011}
Michael Försth and Arne Roos.
\newblock Absorptivity and its dependence on heat source temperature and degree
  of thermal breakdown.
\newblock {\em Fire and Materials}, 35:285–301, 2011.

\bibitem{Hallman_Welker_Sliepcevich_1974}
J.~R. Hallman, J.~R. Welker, and C.~M. Sliepcevich.
\newblock Polymer surface reflectance‐absorptance characteristics.
\newblock {\em Polymer Engineering \& Science}, 14(10):717–723, 1974.

\bibitem{Hood_1966}
James~Dennis Hood.
\newblock {\em A method for the determination of the radiative properties of
  flames}.
\newblock PhD thesis, The University of Oklahoma, 1966.

\bibitem{Boulet_Brissinger_Collin_Acem_Parent_2015}
Pascal Boulet, Damien Brissinger, Anthony Collin, Zoubir Acem, and Gilles
  Parent.
\newblock On the influence of the sample absorptivity when studying the thermal
  degradation of materials.
\newblock {\em Materials}, 8(8):5398–5413, 2015.

\bibitem{Apte2006}
Vivek~B. Apte and Minerals Institute~of Materials.
\newblock {\em Flammability testing of materials used in construction,
  transport and mining}.
\newblock Woodhead Pub., 2006.

\bibitem{Shannon_Butler_2003}
K.S. Shannon and B.W. Butler.
\newblock A review of error associated with thermocouple temperature
  measurement in fire environments.
\newblock {\em International Wildland Fire Ecology}, page 7–9, 2003.

\bibitem{Tranchard_2015}
Pauline Tranchard.
\newblock {\em Modelling the behaviour of a carbon / epoxy composite submitted
  to fire}.
\newblock PhD thesis, L’UNIVERSITÉ DE LILLE 1 SCIENCES ET TECHNOLOGIES,
  2015.

\bibitem{Acem_Brissinger_Collin_Parent_Boulet_Quach_Batiot_Richard_Rogaume_2017}
Zoubir Acem, Damien Brissinger, Anthony Collin, Gilles Parent, Pascal Boulet,
  Thi Hay~Yen Quach, Benjamin Batiot, Franck Richard, and Thomas Rogaume.
\newblock Surface temperature of carbon composite samples during thermal
  degradation.
\newblock {\em International Journal of Thermal Sciences}, 112, Feb 2017.

\bibitem{Sanchez-Carballido2018}
Sergio Sánchez-Carballido, Celeste Justo-María, Juan Meléndez, and Fernando
  López.
\newblock Measurement of the thermal parameters of composite materials during
  fire tests with quantitative infrared imaging.
\newblock {\em Fire Technology}, 54(1):313–333, 2018.

\bibitem{Seggewiss_2011}
Peter Georg~Bernhard Seggewiss.
\newblock {\em Experimentelle und numerische Untersuchung des Verhaltens
  kohlenstofffaserverstärkter Epoxidmatrix-Systeme unter einseitiger
  thermischer und gleichzeitiger mechanischer Belastung}.
\newblock PhD thesis, Universität der Bundeswehr München, 2011.

\bibitem{Melendez2010}
J.~Meléndez, A.~Foronda, J.~M. Aranda, F.~López, and F.~J. López Del~Cerro.
\newblock Infrared thermography of solid surfaces in a fire.
\newblock {\em Measurement Science and Technology}, 21(10), 2010.

\bibitem{Rippe_Lattimer_2015}
Christian~M. Rippe and Brian~Y. Lattimer.
\newblock Full-field surface heat flux measurement using non-intrusive infrared
  thermography.
\newblock {\em Fire Safety Journal}, 78:238–250, 2015.

\bibitem{Bearinger_Hodges_Yang_Rippe_Lattimer_2020}
Elias~D. Bearinger, Jonathan~L. Hodges, Fengchang Yang, Christian~M. Rippe, and
  Brian~Y. Lattimer.
\newblock Localized heat transfer from firebrands to surfaces.
\newblock {\em Fire Safety Journal}, page 103037, January 2020.

\bibitem{Reynolds_1964}
P.~M. Reynolds.
\newblock A review of multicolour pyrometry for temperatures below 1500°c.
\newblock {\em British Journal of Applied Physics}, 15(5):579–589, 1964.

\bibitem{Sanchez-Carballido2017}
Sergio Sánchez-Carballido, Celeste Justo-María, Juan Meléndez, and Fernando
  López.
\newblock A quantitative infrared imaging system for in situ characterization
  of composite materials in fire tests.
\newblock {\em Fire Technology}, 53(3):1309–1331, 2017.

\bibitem{Schuhler_Chaudhary_Vieille_Coppalle_2021}
Eliot Schuhler, Avinash Chaudhary, Benoit Vieille, and Alexis Coppalle.
\newblock {Fire behaviour of composite materials using kerosene burner tests at
  small-scales}.
\newblock {\em Fire Safety Journal}, 121:103290, may 2021.

\bibitem{Li_Gong_Stoliarov_2014}
Jing Li, Junhui Gong, and Stanislav~I. Stoliarov.
\newblock Gasification experiments for pyrolysis model parameterization and
  validation.
\newblock {\em International Journal of Heat and Mass Transfer}, 77:738–744,
  2014.

\bibitem{Reggeti_Agrawal_Bittle_2019}
Shawn~A. Reggeti, Ajay~K. Agrawal, and Joshua~A. Bittle.
\newblock Two-color pyrometry system to eliminate optical errors for spatially
  resolved measurements in flames.
\newblock {\em Applied Optics}, 58(32):8905, 2019.

\bibitem{Zhang_Dai_Lu_Wu_2016}
H.~Zhao and N.~Ladommatos.
\newblock Effects of temperature on the spectral emissivity of c/sic
  composites.
\newblock {\em Progress in Energy and Combustion Science}, 60(3):221–255,
  1998.

\bibitem{Yu_Bauer_Huber_Will_Cai_2021}
Tao Yu, Florian~J. Bauer, Franz~J. Huber, Stefan Will, and Weiwei Cai.
\newblock 4d temperature measurements using tomographic two-color pyrometry.
\newblock {\em Optics Express}, 29(4):5304, 2021.

\bibitem{Arakawa_Saito_Gruver_1993}
A.~Arakawa, K.~Saito, and W.A. Gruver.
\newblock Automated infrared imaging temperature measurement with application
  to upward flame spread studies. part i.
\newblock {\em Combustion and Flame}, 92(3):222--IN2, Feb 1993.

\bibitem{Chan_Fattah_2019}
Q.~N. Chan, I.~M.Rizwanul Fattah, G.~Zhai, H.~L. Yip, T.~B.Y. Chen, A.~C.Y.
  Yuen, W.~Yang, A.~Wehrfritz, X.~Dong, S.~Kook, and et~al.
\newblock Color-ratio pyrometry methods for flame–wall impingement study.
\newblock {\em Journal of the Energy Institute}, 92(6):1968–1976, 2019.

\bibitem{Aphale_DesJardin_2019}
Siddhant~S. Aphale and Paul~E. DesJardin.
\newblock Development of a non-intrusive radiative heat flux measurement for
  upward flame spread using dslr camera based two-color pyrometry.
\newblock {\em Combustion and Flame}, 210:262–278, 2019.

\bibitem{Hunter_Allemand_Eagar_1986}
Gordon~B. Hunter, Charly~D. Allemand, and Thomas~W. Eagar.
\newblock Prototype device for multiwavelength pyrometry.
\newblock {\em Optical Engineering}, 25(11), 1986.

\bibitem{Hunter1984}
B~Hunter, Gordon, Charly~D Allemand, and Thomas~W Eagar.
\newblock An improved method of multi-wavelength pyrometry.
\newblock {\em SPIE Thermosense VII}, 520, 1984.

\bibitem{Inagaki_Okamoto_Fan_1994}
Terumi Inagaki, Yoshizo Okamoto, and Zuofen Fan.
\newblock Temperature measurement and accuracy of bi—colored radiometer
  applying pseudo gray-body approximation.
\newblock In John~R. Snell, Jr., editor, {\em Proceedings Volume 2245,
  Thermosense XVI: An International Conference on Thermal Sensing and Imaging
  Diagnostic Applications}, volume 2245, page 274–285, Mar 1994.

\bibitem{ac20135}
{Federal Aviation Administration}.
\newblock Ac 20-135 - powerplant installation and propulsion system component
  fire protection test methods, standards and criteria with change 1.
\newblock Standard Advisory Circular AC 20-135 with Change 1, International
  Organization for Standardization, US, 2018.

\bibitem{iso2685}
{ISO Central Secretary}.
\newblock Aircraft — environmental test procedure for airborne equipment —
  resistance to fire in designated fire zones.
\newblock Standard ISO 2685:1998, International Organization for
  Standardization, Geneva, CH, 1998.

\bibitem{Beland_Sc_2018}
Mathieu Béland.
\newblock {\em Study and Design of a Small Kerosene Burner}.
\newblock Master thesis, Universit\'e Laval, 2018.

\bibitem{FAA}
E.~P. Burke and T.~G. Horeff.
\newblock {Power plant engineering report No. 3A Standard fire test apparatus
  and procedure (for flexible hose assemblies)}.
\newblock Technical report, U.S. Department of transportation; Federal aviation
  administation, 3 1978.

\bibitem{Ochs_2008}
Robert Ian (Federal Aviation~Administration) Ochs.
\newblock {Development of a Next-Generation Burner for Testing Thermal Acoustic
  Insulation Burnthrough Resistance -- DOT/FAA/AR-TN09/23}.
\newblock Technical Report May, U.S. Department of Transportation Federal
  Aviation Administration, Atlantic City, New Jersey, USA, 2009.

\bibitem{Kao_Tambe_Ochs_Summer_Jeng_2017}
Yi~Huan Kao, Samir~B. Tambe, Robert Ochs, Steve Summer, and San~Mou Jeng.
\newblock Experimental study of the burner for faa fire test: Nexgen burner.
\newblock {\em Fire and Materials}, 41(7):898–907, 2017.

\bibitem{Gibson_Torres_Browne_Feih_Mouritz_2010}
A.~G. Gibson, M.~E.Otheguy Torres, T.~N.A. Browne, S.~Feih, and A.~P. Mouritz.
\newblock High temperature and fire behaviour of continuous glass
  fibre/polypropylene laminates.
\newblock {\em Composites Part A: Applied Science and Manufacturing},
  41(9):1219–1231, 2010.

\bibitem{Horold_Schartel_Trappe_Korzen_Naumann_2012}
A.~Hörold, B.~Schartel, V.~Trappe, M.~Korzen, and M.~Naumann.
\newblock Structural integrity in fire: An intermediate-scale approach.
\newblock {\em ECCM 2012 - Composites at Venice, Proceedings of the 15th
  European Conference on Composite Materials}, page 24–28, June 2012.

\bibitem{Hijazi2011}
A.~Hijazi, S.~Sachidanandan, R.~Singh, and V.~Madhavan.
\newblock {A calibrated dual-wavelength infrared thermometry approach with
  non-greybody compensation for machining temperature measurements}.
\newblock {\em Measurement Science and Technology}, 22(2), 2011.

\bibitem{Jo2017}
Hang~Jin Jo, Jonathan~L. King, Kyle Blomstrand, and Kumar Sridharan.
\newblock {Spectral emissivity of oxidized and roughened metal surfaces}.
\newblock {\em International Journal of Heat and Mass Transfer},
  115:1065--1071, 2017.

\bibitem{Ballester2010}
Javier Ballester and Tatiana Garc{\'{i}}a-Armingol.
\newblock {Diagnostic techniques for the monitoring and control of practical
  flames}.
\newblock {\em Progress in Energy and Combustion Science}, 36(4):375--411, aug
  2010.

\bibitem{Savino2017}
L.~Savino, M.~{De Cesare}, M.~Musto, G.~Rotondo, F.~{De Filippis}, A.~{Del
  Vecchio}, and F.~Russo.
\newblock {Free emissivity temperature investigations by dual color applied
  physics methodology in the mid- and long-infrared ranges}.
\newblock {\em International Journal of Thermal Sciences}, 117:328--341, 2017.

\bibitem{Vollmer_Mollmann_2018}
Michael Vollmer and Klaus-Peter M{\"{o}}llmann.
\newblock {\em {Infrared Thermal Imaging}}.
\newblock Wiley-VCH Verlag GmbH \& Co. KGaA, Weinheim, Germany, second edi
  edition, dec 2017.

\bibitem{Swanson2018}
D.~Swanson and J.~Wolfrum.
\newblock {Time to failure modeling of carbon fiber reinforced polymer
  composites subject to simultaneous tension and one-sided heat flux}.
\newblock {\em Journal of Composite Materials}, 52(18), 2018.

\bibitem{Saunders_2001}
Peter Saunders.
\newblock On the effects of temperature dependence of spectral emissivity in
  industrial radiation thermometry.
\newblock {\em High Temperatures - High Pressures}, 33(5):599–610, 2001.

\bibitem{Jones_Mason_Williams_2019}
J.M. Jones, P.E. Mason, and A.~Williams.
\newblock A compilation of data on the radiant emissivity of some materials at
  high temperatures.
\newblock {\em Journal of the Energy Institute}, 92(3):523–534, Jun 2019.

\bibitem{Li_Strieder_2009}
Xiangning Li and William Strieder.
\newblock Emissivity of high-temperature fiber composites.
\newblock {\em Industrial and Engineering Chemistry Research},
  48(4):2236–2244, 2009.

\bibitem{Athanasopoulos2012}
N~Athanasopoulos, D~Sikoutris, T~Panidis, and V~Kostopoulos.
\newblock Numerical investigation and experimental verification of the joule
  heating effect of polyacrylonitrile-based carbon fiber tows under high vacuum
  conditions.
\newblock {\em Journal of Composite Materials}, 46(18):2153–2165, Aug 2012.

\bibitem{Kim_Dembsey_2015}
E.~Kim and N.~Dembsey.
\newblock Parameter estimation for comprehensive pyrolysis modeling: Guidance
  and critical observations.
\newblock {\em Fire Technology}, 51(2), 2015.

\bibitem{Berardi_Dembsey_2015}
Umberto Berardi and Nicholas Dembsey.
\newblock Thermal and fire characteristics of frp composites for architectural
  applications.
\newblock {\em Polymers}, 7(11):2276–2289, 2015.

\bibitem{Kreith2004}
Frank Kreith, Robert~F. Boehm, George~D. Raithby, K.~G.T. Hollands, N.~V.
  Suryanarayana, Michael~F. Modest, Van~P. Carey, John~C. Chen, Noam Lior,
  Ramesh~K. Shah, and et~al.
\newblock Heat and mass transfer.
\newblock {\em The CRC Handbook of Mechanical Engineering, Second Edition},
  pages 4--1--4–357, 2004.

\bibitem{Gordon2012}
Andrew~J. Gordon, Kyle~L. Walton, Tushar~K. Ghosh, Sudarshan~K. Loyalka,
  Dabir~S. Viswanath, and Robert~V. Tompson.
\newblock {Hemispherical total emissivity of Hastelloy N with different surface
  conditions}.
\newblock {\em Journal of Nuclear Materials}, 426(1-3):85--95, jul 2012.

\bibitem{Evans1925}
Ulick~R. Evans.
\newblock {The colours due to thin films on metals}.
\newblock {\em Proceedings of the Royal Society of London. Series A, Containing
  Papers of a Mathematical and Physical Character}, 107(742):228--237, feb
  1925.

\bibitem{Higginson2015}
R.~L. Higginson, C.~P. Jackson, E.~L. Murrell, P.~A.Z. Exworthy, R.~J.
  Mortimer, D.~R. Worrall, and G.~D. Wilcox.
\newblock {Effect of thermally grown oxides on colour development of stainless
  steel}.
\newblock {\em Materials at High Temperatures}, 32(1-2):113--117, 2015.

\bibitem{Roebuck2013}
B.~Roebuck, G.~Edwards, and M.~G. Gee.
\newblock {Characterisation of oxidising metal surfaces with a two colour
  pyrometer}.
\newblock {\em http://dx.doi.org/10.1179/174328405X46169}, 21(7):831--840, jul
  2013.

\bibitem{Gnessougou2019}
Serge-olivier Gnessougou, Alexandrine Huot, and Martin Larivi{\`{e}}re-bastien.
\newblock {Temperature Calculation of Non-Burning \& Burning Materials Exposed
  to a Flame Using a Multispectral Infrared Camera}.
\newblock In {\em 11th U. S. National Combustion Meeting Organized by the
  Western States Section of the Combustion Institute March 24–27, 2019
  Pasadena, California}, pages 1--8, 2019.

\bibitem{Petrov_Reznik_1972}
V.~A. Petrov and V.~Yu Reznik.
\newblock Measurement of the emissivity of quartz glass.
\newblock {\em High Temperatures - High Pressures}, 4:687–693, 1972.

\bibitem{Barnes_Forsythe_Adams_1947}
B.~T. Barnes, W.~E. Forsythe, and E.~Q. Adams.
\newblock The total emissivity of various materials at 100–500°c.
\newblock {\em Journal of the Optical Society of America}, 37(10):804, Oct
  1947.

\end{thebibliography}

\end{document}